\pdfoutput=1 
\documentclass[a4paper, 11pt]{article}
\usepackage[dvipsnames]{xcolor}

\usepackage{jheppub} 

\makeatletter  
\gdef\@fpheader{}  
\makeatother 

\usepackage[T1]{fontenc} 
\usepackage{tangocolors}
\usepackage{tikz}
\usepackage{tikz-3dplot}
\usetikzlibrary{arrows, decorations,backgrounds, patterns}
\usetikzlibrary{decorations.pathreplacing ,decorations.markings}
\usepackage{amssymb}
\usetikzlibrary{decorations.pathmorphing,backgrounds,shapes,arrows,shadows}
\usetikzlibrary{patterns.meta}
\usetikzlibrary{patterns,decorations.pathmorphing}
\tikzset{
    snake it/.style={decorate, decoration=snake}
}
\usepackage{pgfplots}
\pgfplotsset{compat=1.11}
\usepgfplotslibrary{fillbetween}
\usetikzlibrary{intersections}
\pgfdeclarelayer{bg}
\pgfsetlayers{bg,main}
\tikzset{zigzag/.style={decorate,decoration=zigzag}}
\tikzset{snake it/.style={decorate, decoration=snake}}
\makeatletter
\def\@hex@@Hex#1%
 {\if a#1A\else \if b#1B\else \if c#1C\else \if d#1D\else
  \if e#1E\else \if f#1F\else #1\fi\fi\fi\fi\fi\fi \@hex@Hex}
\makeatother

\usepackage[all]{xy}
\usepackage[percent]{overpic}
\usepackage{slashed}
\usepackage{wrapfig}
\usepackage{tabu}
\usepackage{diagbox}
\usepackage{mathrsfs,amsmath,amssymb,amsthm,amsfonts,tikz,graphicx,accents,hyperref, color}
\usepackage{dsfont,epiolmec, latexsym, stmaryrd, comment}
\usepackage{slashed,ccaption}
\usepackage{mathrsfs, calligra}
\usepackage{leftidx}
\usepackage{import}
\usepackage{multirow}
\usepackage{amsfonts}
\usepackage{pifont}
\usepackage{tabularx}
\usepackage{cancel}
\usepackage[utf8]{inputenc}
\usetikzlibrary{intersections,calc}
\usepackage{ifthen}
\usepackage{amsmath}
\usepackage{physics}
\usepackage{cancel}
\usepackage{caption} 
\usepackage{subcaption}

\usetikzlibrary{patterns.meta}
\usepackage{array}
\usepackage{bm}

\hypersetup{ linktoc=all,
    colorlinks, linkcolor={palatinateblue},
    citecolor={brightpink}, urlcolor={amaranth}
}

\graphicspath{{Images/}}

\renewcommand{\d}[1]{\ensuremath{\operatorname{d}\!{#1}}}

\def\sideremark#1{\ifvmode\leavevmode\fi\vadjust{\vbox to0pt{\vss
 \hbox to 0pt{\hskip\hsize\hskip1em
 \vbox{\hsize2cm\tiny\raggedright\pretolerance10000
 \noindent #1\hfill}\hss}\vbox to8pt{\vfil}\vss}}}%
\DeclareSymbolFont{extraup}{U}{zavm}{m}{n}
\DeclareMathSymbol{\varheart}{\mathalpha}{extraup}{86}
\DeclareMathSymbol{\vardiamond}{\mathalpha}{extraup}{87}
\makeatletter
\renewcommand*{\@fnsymbol}[1]{\ensuremath{\ifcase#1\or \clubsuit \or \vardiamond \or \varheart\or
    \spadesuit\or \mathparagraph\or \|\or **\or \dagger\dagger
    \or \ddagger\ddagger \else\@ctrerr\fi}}
\makeatother

\definecolor{rosy}{RGB}{230,235,252}
\definecolor{myframetitle}{RGB}{90,89,170}
\definecolor{myblocktitle}{RGB}{140,185,249}
\definecolor{mytitle}{RGB}{10,80,26}

\definecolor{darkgreen}{RGB}{27,130,45}
\definecolor{darkblue}{rgb}{0,0,0.3}
\definecolor{darkred}{rgb}{0.7,0,0}

\definecolor{light gray}{RGB}{220,220,220}
\definecolor{dark purple}{RGB}{108,0,217}
\definecolor{pink}{RGB}{190,20,100}
\definecolor{orang}{RGB}{193,63,0}
\definecolor{green}{RGB}{11,98,17}
\definecolor{darkpink}{RGB}{153,0,76}
\definecolor{bluegreen}{RGB}{0,102,102}
\definecolor{greenlagan}{RGB}{0,102,0}
\definecolor{redgreen}{RGB}{102,102,0}
\definecolor{Redgreen}{RGB}{153,76,0}
\definecolor{vividviolet}{rgb}{0.62, 0.0, 1.0}
\definecolor{amaranth}{rgb}{0.9, 0.17, 0.31}
\definecolor{palatinateblue}{rgb}{0.15, 0.23, 0.89}
\definecolor{brightpink}{rgb}{1.0, 0.0, 0.5}
\definecolor{cornflowerblue}{rgb}{0.39, 0.58, 0.93}
\definecolor{deepcarminepink}{rgb}{0.94, 0.19, 0.22}
\definecolor{radicalred}{rgb}{1.0, 0.21, 0.37}

\newcommand\mgnote[1]{\textcolor{magenta}{\bf [Mahdi:\,#1]}}

\usepackage[most]{tcolorbox}

\tcbset{highlight math style={left=02mm,right=02mm,top=02mm,bottom=02mm}} 
\usepackage{empheq}

\DeclareFontFamily{OT1}{rsfs}{}

\DeclareFontShape{OT1}{rsfs}{m}{n}{ <-7> rsfs5 <7-10> rsfs7 <10->rsfs10}{} 

\DeclareMathAlphabet{\mycal}{OT1}{rsfs}{m}{n}

\newcommand{\be}{\begin{equation}}
\newcommand{\ee}{\end{equation}}
\newcommand{\bea}{\begin{eqnarray}}
\newcommand{\eea}{\end{eqnarray}}
\makeatletter \@addtoreset{equation}{section}

\newcommand\tcb{\textcolor{blue}}

\newcommand{\bdry}{\mathcal{B}}
\newcommand{\cauchy}{\mathcal{C}}

\newcommand{\Bplus}{\bdry^{^{\scriptscriptstyle >}}}
\newcommand{\MCL}{\mathcal{M}_\mathcal{C}^{^{\scriptscriptstyle >}}}
\newcommand{\MCS}{\mathcal{M}_\mathcal{C}^{^{\scriptscriptstyle <}}}
\newcommand{\MBL}{\mathcal{M}_\mathcal{B}^{^{\scriptscriptstyle >}}}
\newcommand{\MBS}{\mathcal{M}_\mathcal{B}^{^{\scriptscriptstyle <}}}

\begin{document}

\newcommand{\mytitle}{\begin{center}{\Large{\textbf{Covariant Phase Space Formalism for Fluctuating Boundaries
}}}
\end{center}}

\title{{\mytitle}}
\author[a]{H.~Adami}
\author[b,c]{, M.~Golshani}
\author[c,d]{, M.M.~Sheikh-Jabbari}
\author[b,c]{, V.~Taghiloo}
\author[b,c]{, M.H.~Vahidinia}

\affiliation{$^a$ Shanghai Institute for Mathematics and Interdisciplinary Sciences (SIMIS), Shanghai, 200433, China}
\affiliation{$^b$ Department of Physics, Institute for Advanced Studies in Basic Sciences (IASBS),\\ 
P.O. Box 45137-66731, Zanjan, Iran}
\affiliation{$^c$ School of Physics, Institute for Research in Fundamental
Sciences (IPM),\\ P.O.Box 19395-5531, Tehran, Iran}
\affiliation{$^d$ The Abdus Salam ICTP, Strada Costiera 11,  I-34151 Trieste, Italy}
\emailAdd{
hadami@simis.cn, mahdig@iasbs.ac.ir, 
jabbari@theory.ipm.ac.ir, v.taghiloo@iasbs.ac.ir, vahidinia@iasbs.ac.ir
}
\abstract{We reconsider formulating $D$ dimensional gauge theories, with the focus on the case of gravity theories, in spacetimes with boundaries. We extend covariant phase space formalism to the cases in which boundaries are allowed to fluctuate. We analyze the symplectic form, the freedoms (ambiguities), and its conservation for this case. We show that boundary fluctuations render all the surface charges integrable. We study the algebra of charges and its central extensions, charge conservation, and fluxes. We briefly comment on memory effects and questions regarding semiclassical aspects of black holes in the fluctuating boundary setup.}
\maketitle
\section{Introduction}\label{sec:Intro}

In physics, we typically deal with a Cauchy problem, that dynamics of a system is (uniquely) determined by the set of initial data we provide on a constant time slice, a (partial) Cauchy surface ${\cal C}$. However,  spacetimes over which we want to formulate a field theory generically also have a boundary, a codimension-one timelike surface ${\cal B}$; e.g. Maxwell theory in a box or field theories on an anti-de Sitter (AdS) space. In these cases, we deal with a boundary$+$initial value problem where besides the initial data we should also specify information on the boundary to determine the dynamics. 

Noether's theorem relates symmetries to (conserved) charges which are integrals of the Noether current over constant time slices ${\cal C}$. In theories with local symmetries these integrals by virtue of equations of motion reduce to a surface integral, an integral over codimension-two surfaces ${\cal S}$ \cite{Lee:1990nz, Henneaux:1994lbw}. In our setup, ${\cal S}$ is the intersection of boundary ${\cal B}$ and (partial) Cauchy surface ${\cal C}$ as depicted in Fig. \ref{fig:3dsetup}. 

In gauge or diffeomorphism invariant theories, the role of local symmetries is to remove unphysical (non-dynamical) bulk degrees of freedom (d.o.f), redundancies added for having a covariant formulation of the theory. In the presence of codimension-one boundaries (including the partial Cauchy surface) local symmetries should be studied more carefully. It is known that for a consistent formulation of the theory, one should add boundary d.o.f to guarantee gauge/diffeomorphism invariance of the theory at the boundaries. Each physics problem is defined through its appropriate boundary conditions which in turn impose restrictions on the form of local gauge/diffeomorphism transformations at the boundary, and importantly, requires a specific set of boundary d.o.f with specific dynamics. From a different viewpoint, the presence of boundary conditions imposes restrictions on the gauge/diffeomorphisms at the boundary. This in turn, renders some of the gauge transformations on codimension-one surfaces (boundaries) physical, i.e. some physically inequivalent configurations (that differ by the associated configuration of boundary d.o.f) are mapped to each by physical ``improper'' \cite{Henneaux:1994lbw, Benguria:1976in} gauge/diffeomorphism transformations. It is known that there are surface charges (integrals over ${\cal S}$) associated with such physical gauge/diffeomorphisms and that these charges can be used to label boundary d.o.f \cite{Strominger:2017zoo}.
 
There are several formulations to study surface charges in gauge and gravity theories \cite{Henneaux:1994lbw, Szabados:2009eka, crnkovic1987covariant, Lee:1990nz}. Among them, we are interested in Covariant Phase Space Formalism (CPSF) \cite{crnkovic1987covariant, Lee:1990nz, Iyer:1994ys, Wald:1999wa}, see \cite{Seraj:2016cym,  Compere:2018aar, Grumiller:2022qhx} for reviews. While our analyses and main results apply to any theory with a local symmetry, we will be mainly interested in gravity theories, where one deals with diffeomorphisms as local symmetry generators. In this context, the question of boundary symmetries, currents, and importantly a generally invariant (or covariant) notion of charges is still not fully settled. The cases of usual interest are the  asymptotic null boundary of flat space \cite{Sachs:1962wk, Sachs:1962zza, Bondi:1962px, Ashtekar:1981bq, Ashtekar:1978zz} and \cite{Barnich:2009se, Barnich:2010eb, Barnich:2010ojg, Barnich:2011mi,Barnich:2013axa, Chen:2021kug,  Chandrasekaran:2020wwn, Geiller:2022vto,  Donnay:2023mrd, Perez:2022jpr,Flanagan:2019vbl, Laddha:2020kvp, Geiller:2024amx,  Krishnan:2023zdd, Rignon-Bret:2024gcx,Rignon-Bret:2024wlu, Ashtekar:2024mme, McNees:2023tus}, or a null boundary in non-asymptotic region as   horizon of  a black hole  \cite{Donnay:2015abr,Donnay:2016ejv,Grumiller:2019fmp, Chandrasekaran:2018aop, Grumiller:2020vvv,Adami:2020amw,Adami:2021nnf, Adami:2020ugu, Adami:2021sko, Chandrasekaran:2021hxc,Chandrasekaran:2023vzb,Freidel:2022bai, Donnay:2020yxw, Freidel:2022vjq, Sheikh-Jabbari:2022mqi, Odak:2023pga, Ciambelli:2023mir, Mao:2022ldv}; or cases with timelike boundaries that appear in causal boundary of AdS spaces \cite{Brown:1986nw, Ashtekar:1984zz} and \cite{Mao:2019ahc, Compere:2020lrt, Fiorucci:2020xto, Alessio:2020ioh,Adami:2022ktn}.

CPSF, as the name suggests, has the advantage of providing a covariant formulation for the computation of surface charges. However, it comes with a caveat: Unlike the usual Noether charge, CPSF yields charge variations (over the field or solution space) and there always remains the question of whether these charge variations are integrable and whether one can define charges \cite{Lee:1990nz}. Besides (non)integrability of charges, one may also wonder about their conservation; while the two notions of integrability and conservation are conceptually distinct, they are closely related \cite{Seraj:2016cym, Adami:2020amw}. In general, a charge variation may be viewed as a sum of an integrable part and a ``flux'' \cite{Barnich:2010eb}. It has been argued that integrability of the charge may depend on the slicing of the solution space and that charges can become integrable in suitable slicings, provided that the flux is not a ``genuine one'' \cite{Adami:2020ugu, Adami:2021nnf}.\footnote{A genuine flux is by definition only made out of bulk modes, such that in the absence of flux of bulk modes through the boundary the genuine flux vanishes \cite{Adami:2021nnf}. Note also that the algebra of charges as well as the central charges all depend on the slicing of the solution space and are not inherent characteristics of the solution space \cite{Adami:2020ugu}.}

Motivated by the fact that surface charges are given by codimension-two integrals, it has been argued that an appropriate setup can be focusing on the codimension-two surface ${\cal S}$, instead of considering the problem on a given boundary ${\cal B}$ and/or  Cauchy surface ${\cal C}$. This codimension-two surface which is the intersection of ${\cal B}$ and ${\cal C}$, is usually called the corner. The significance of the corner viewpoint in the context of gravity has been already noted and emphasized in the seminal Brown-York ``quasi-local'' charges \cite{Brown:1992br}. If the surface charges are well-defined and integrable, they can be used to label boundary d.o.f. This has led to a renewed attention to the corner viewpoint \cite{Donnelly:2016auv, Speranza:2017gxd, Freidel:2020xyx, Freidel:2020svx, Freidel:2020ayo, Freidel:2021cjp, Ciambelli:2022vot, Ciambelli:2021nmv,  Freidel:2021dxw, Carrozza:2022xut, Freidel:2023bnj}. 

In this work, we revisit the important issue of charge integrability in the corner viewpoint. This question has been discussed in \cite{Ciambelli:2021nmv, Freidel:2021dxw, Speranza:2022lxr, Carrozza:2022xut}. While our main results are in agreement with these earlier work, we give a new viewpoint and shed a new light on the problem. We show that if we allow generic fluctuations of the Cauchy surface ${\cal C}$ and the boundary ${\cal B}$, all the surface charges  become integrable and the integrable charge is essentially the Noether charge. To achieve this we carefully analyze the on-shell symplectic form with special attention paid to the boundary fluctuations and corner terms.

\paragraph{Outline of the paper.} In section \ref{sec:setup}, we lay the grounds by introducing the essential mathematical framework. In section \ref{sec:CFSP}, we delve into the development of CPSF for fluctuating boundaries. This section is devoted to analyses of symplectic two-form, study the freedoms (ambiguities) in its determination, and discuss its conservation. In section \ref{sec:surface-charge}, we study surface charges and establish that in a fluctuating boundary setup, they are always integrable. We study charge algebra and show how the central charges arise in the algebra. We also discuss charge conservation and the freedoms in the definition of the charge.  The final section \ref{sec:conc}, is dedicated to summary, discussion, and outlook. In three appendices we have gathered useful equations, identities, and details of computations. 
\section{Mathematical setup}\label{sec:setup}

Consider a $D$ dimensional spacetime $\MBL$ spanned by coordinates $x^\mu$ that possesses a codimension-one boundary ${\cal B}$. While we take this boundary to be timelike, without loss of generality, it can also be taken to be spacelike or null. To facilitate the analysis, we view $\MBL$ as part of a hypothetical boundary-less manifold $\mathcal{M}$.  The boundary of $\MBL$ is a codimension-one timelike hypersurface $\mathcal{B}(x)=0$, where $\mathcal{B}(x)$ is a generic function of spacetime.\footnote{Without loss of generality one can assume the boundary to be given by $\mathcal{B}(x^\mu)=r-B(t,x^A)$, where $x^{\mu}=(t,r,x^A)$ and $(t,x^A)$ are the coordinates on $\bdry$.} The presence of this hypersurface splits the manifold $\mathcal{M}$ into two regions: $\MBS:=\{x^{\mu}\, \vert\, \bdry(x)\leq 0\}$ and $\MBL:=\{x^{\mu}\, \vert \, \bdry(x) \geq 0\}$.  
One may foliate spacetime into constant $\bdry(x)$ timelike hypersurfaces.  

Analogously, we introduce a spacelike codimension-one hypersurface $\mathcal{C}(x)=0$, which we assume to be a (partial) Cauchy surface denoted by ${\cal C}$.  ${\cal C}$ splits spacetime into two regions:  $\MCS:=\{x^{\mu}\, |\, \cauchy(x)\leq0\}$ and $\MCL:=\{x^\mu\, |\, \cauchy(x) \geq 0\}$. Constant ${\cal C}(x)$ surfaces foliate spacetime into spacelike hypersurfaces.  

\begin{figure}
    \centering
    \includegraphics[scale=.6]{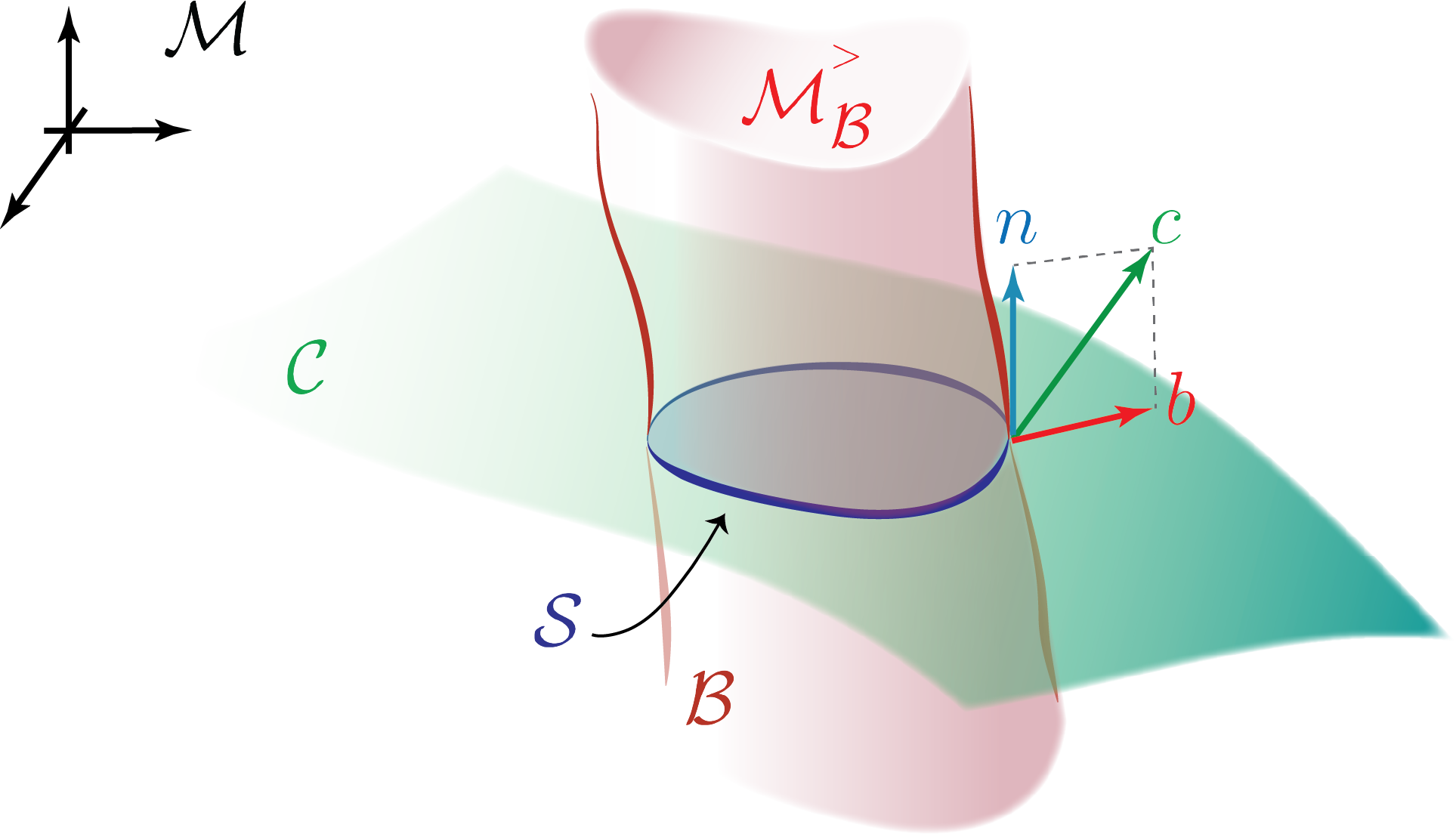}
    \caption{The illustration depicts the $D$-dim. manifold $\textcolor{red}{\MBL}$ (presented in red), which is a submanifold of a $D$-dim. boundary-less manifold $\mathcal{M}$. The boundary of $\textcolor{red}{\MBL}$ is represented by the codimension-one surface $\textcolor{BrickRed}{\bdry}$ (illustrated in dark red lines). Codimension-one spacelike hypersurface $\textcolor{Green}{\cauchy}$ shows a Cauchy surface (displayed in green) and its intersection with $\textcolor{red}{\MBL}$ is a codimension-two hypersurface $\textcolor{Blue}{\mathcal{S}}$ (shown in dark blue). $\textcolor{red}{b}$ and $\textcolor{Green}{c}$ are normals to $\textcolor{BrickRed}{\bdry}$ and $\textcolor{Green}{\cauchy}$ respectively  that are not necessarily orthogonal to each other. The vector $\textcolor{NavyBlue}{n}$ is timelike and satisfies the condition $b_{\mu}n^{\mu}=0$.}
    \label{fig:3dsetup}
\end{figure}

The two  hypersurfaces $\mathcal{B}(x)=0$ and $\mathcal{C}(x)=0$ intersect in a codimension-two surface $\mathcal{S}$. 
We take ${\cal S}$ to be a compact and smooth surface. Moreover, we take all functions in our calculations to be smooth over ${\cal S}$ and will hence drop integrals of total divergences on ${\cal S}$. It should be noted that the presence of ${\cal S}$ further splits the boundary of $\MBL$, denoted by $\bdry$, 
into future and past codimension-one regions. In the rest of this paper, we are only interested in the future part that is denoted by $ 	\Bplus:= 	\bdry \cap \MCL$. These have been summarized as follows and depicted in Fig. \ref{fig:3dsetup},
 \begin{equation}    
\begin{split}	
\mathcal{M}_\mathcal{B}^{^{\scriptscriptstyle >}}:=\qty{x^\mu\vert \mathcal{B}(x)\geq 0}\, , & \qquad \mathcal{M}_\mathcal{B}^{^{\scriptscriptstyle <}}:=\qty{x^\mu\vert \mathcal{B}(x)\leq  0}\, ,\\
\MCL:=\qty{x^\mu\vert \mathcal{C}(x)\geq 0}\, , & \qquad  \MCS:=\qty{x^\mu\vert \mathcal{C}(x)\leq 0}\, ,\\ 
 {\cal C}=\mathcal{M}_\mathcal{C}^{^{\scriptscriptstyle >}} \cap \mathcal{M}_\mathcal{C}^{^{\scriptscriptstyle <}}\, , & \qquad
{\cal B}=\mathcal{M}_\mathcal{B}^{^{\scriptscriptstyle >}} \cap \mathcal{M}_\mathcal{B}^{^{\scriptscriptstyle <}}\, ,\\
\Bplus:= 	\mathcal{B}\cap \mathcal{M}_{\mathcal{C}}^{^{\scriptscriptstyle >}}&:=\qty{x^\mu\vert \mathcal{B}(x)= 0 \, , \mathcal{C}(x)\geq 0 }\, ,\\ 
\mathcal{M}=\mathcal{M}_\mathcal{C}^{^{\scriptscriptstyle >}} \cup \mathcal{M}_\mathcal{C}^{^{\scriptscriptstyle <}}=\MBL \cup \MBS\, , & \qquad  \mathcal{S}:=\qty{x^\mu \vert \mathcal{C}=0,\mathcal{B}=0 }= {\cal B}\cap {\cal C}\, .
\end{split}
\end{equation}

\paragraph{Geometric quantities.} One can define some  geometric quantities associated with ${\cal B}$ and ${\cal C}$ codimension-one surfaces. Unit vectors normal to constant $\bdry$ and $\cauchy$ hypersurfaces are
\begin{subequations}
    \begin{align}
      &b_\mu  = -\, N_{\bdry} \, \partial_\mu \bdry \, , \qquad N_{\bdry}= (  \partial_\mu\bdry\, \partial^\mu\bdry )^{-1/2}\,\, ,  \qquad b_\mu b^\mu=1, \, \label{bmu-Nb}\\
      &c_\mu  = -\, N_{\cauchy} \, \partial_\mu \cauchy \, , \qquad N_{\cauchy}= ({-}  \partial_\mu\cauchy\, \partial^\mu\cauchy)^{-1/2}\,\, ,  \qquad c_\mu c^\mu=-1 \, ,\label{cmu-Nc}
    \end{align}
\end{subequations}
where we raise and lower spacetime indices by the metric $g_{\mu\nu}$ or its inverse $g^{\mu\nu}$. 
The two one-form fields $b_\mu, c_\mu$ form a basis for the 2-dim. space transverse to the codimension-two corner ${\cal S}$. However, they are not necessarily orthogonal to each other. To have an orthogonal basis, we introduce a timelike covariant vector $n_\mu$ instead of $c_\mu$ as follows
\begin{equation}\label{nmu-N}
  \hspace{-0.4 cm}  n_\mu =- N  (\partial_\mu \cauchy - \mathcal{A} \, \partial_\mu \bdry )\, , \quad N=(N_{\cauchy}^{-2}+\mathcal{A}^2 N_{\bdry}^{-2} )^{-1/2}\, , \quad \mathcal{A} = N_\bdry^2 ( \partial_\mu \bdry \partial^\mu \cauchy) \, , \quad n_\mu n^\mu=-1\, .
\end{equation}
This basis and non-orthonormal basis $(b_\mu, c_\mu)$ are related as
\begin{equation}
    b_\mu n^\mu=0\, ,\qquad c_\mu n^\mu=-\frac{N_{\cauchy}}{N}\, ,\qquad b_\mu c^\mu=\frac{N_{\cauchy}}{N_{\bdry}}{\cal A}\, .
\end{equation} 
It should be noted that in the non-orthonormal basis $(b_\mu, c_\mu)$, the three components of the 2-dim. metric are parametrized by ${\cal A}(x), {\cal B}(x), {\cal C}(x)$. Before moving further, we note that the choice of $b_\mu$ being spacelike and $c_\mu$ or $n_\mu$ being timelike may be relaxed and one can even consider the cases where one or both of these directions are null. However, for definiteness, we will keep the choices made. 

Induced metrics on the boundary $\bdry$ and corner $\mathcal{S}$ are 
respectively defined as
\begin{subequations}
    \begin{align}
        & \gamma_{\mu\nu}:=g_{\mu\nu}- b_{\mu}\, b_{\nu}\, , \label{induced-Sigma}\\
        & q_{\mu\nu}:=g_{\mu\nu}- b_{\mu}\, b_{\nu}+ n_{\mu}\, n_{\nu}\, . \label{induced-Sigmac}
    \end{align}
\end{subequations}
The square root of the determinant of the metrics are related as,  
\begin{equation}
\sqrt{-g}= N_\bdry \, \sqrt{- \gamma}={N}\,  N_\bdry \, \sqrt{q}\, .  
\end{equation}
${\cal E}_{\mu\nu}$, the  bi-normal to  the codimension-two surface $\mathcal{S}$ is defined as
\begin{equation}\label{bi-normal}
        {\cal E}_{\mu\nu}:=2 N^{-1}\,N_{\bdry}^{-1}\,   n_{[\mu} b_{\nu]}=
        2 N_\cauchy^{-1}\, N_\bdry^{-1}\, c_{[\mu}\, b_{\nu]}
        =2 \partial_{[\mu}\cauchy\, \partial_{\nu]}\bdry\, ,\qquad \mathcal{E}_{\mu \nu} \mathcal{E}^{\mu \nu} = -2 N^{-2} N_\bdry^{-2},
\end{equation}
where we use the antisymmetrization convention $c_{[\mu}\, b_{\nu]}=\frac12(c_{\mu}b_{\nu}-c_{\nu}b_{\mu})$.
One can use $b_{\mu}$ and $\mathcal{E}_{\mu \nu}$ to define measure of integrals over $\bdry$ and $\mathcal{S}$:
\begin{equation}\label{notation-int}
\int_{\bdry}\d{}x_{\mu}:=\int_{\bdry}\, \d{}^{D-1}x\, N_{\bdry}^{-1}\, b_{\mu}\, , \qquad \int_{\cal S}\d{}x_{\mu\nu}:=\frac{1}{2}\int_{\cal S}\, \d{}^{D-2}x\, {\cal E}_{\mu\nu}\, .
\end{equation}

\paragraph{$(p,q)$-forms.} Besides the geometric quantities above we also deal with fields on spacetime, generically denoted by $\Phi$. The variations $\delta\Phi$ are hence one-forms on the solution space. We use the notation $(p,q)$-form to denote a $p$-form in spacetime which is a $q$-form in the field space. Hence ${\cal E}_{\mu \nu}$ is a (2,0)-form, $\delta{\cal A}, \delta{\cal B},\delta{\cal C}$ are (0,1)-forms,  $\delta b_{\mu}$ and $\delta c_{\mu}$ are $(1,1)$-forms and $\delta {\cal E}_{\mu \nu}$ is a $(2,1)$-form, and
\begin{equation}\label{basic-form-relation-1}
   \delta{\cal E}_{\mu\nu} 
   =2 \partial_{[\mu} \chi^*_{\nu]}\,,
\end{equation}
where $(1,1)$-form $\chi^{*}_{\mu}$ and its corresponding vector are defined as 
\begin{subequations}\label{chi-chi*-def}
\begin{align}
\chi^*_{\mu}&:=\partial_\mu \bdry\, \delta \cauchy-\partial_\mu \cauchy\, \delta \bdry=\mathcal{E}_{\mu\nu}\chi^{\nu}\, , \label{chi*-def}\\
\chi^{\mu}&:=-N\, {n^{\mu}}\, \delta \cauchy + N_{\bdry}\, {b^{\mu}}\, \delta \bdry+ N\, \mathcal{A}\, n^{\mu}\, \delta \bdry={ N^2 N_\bdry^2 \,{\cal E}^{\mu\nu}\chi^{*}_{\nu}}\, . \label{chi-def}
\end{align}
\end{subequations}
We can also define the $(0,2)$-form $\mathbb{S}$ and the $(1,2)$-form $\partial_\mu \mathbb{S}$ as 
\begin{equation}\label{basic-form-relation-2}
\mathbb{S}:=\delta\cauchy\wedge\delta \bdry=\frac{1}{2}\chi^{\mu}\wedge \chi^*_{\mu}\, , \qquad
    \partial_\mu \mathbb{S} ={-}\delta \chi^*_{\mu}\, ,
\end{equation}
where $\wedge$ denotes the wedge product over field space.\footnote{In \cite{Freidel:2021dxw,Carrozza:2022xut} $\chi$ has been viewed as a Maurer–Cartan form. However, as \eqref{basic-form-relation-2} shows $(1,1)$-form $\chi$, while resembling,  is not strictly speaking a Maurer–Cartan form. } 

We close this section by the comment that \eqref{basic-form-relation-1} and \eqref{basic-form-relation-2} may be viewed as equations for the $(1,1)$-form $\chi$. These equations in the notation of $(p,q)$-forms may be formally written as,\footnote{Note that \eqref{E-S-chi} implies $\delta{\mathbb{S}}=0,  \d{} {\cal E}=0, \d{}\delta\chi^*=\delta\!\d{}\chi^*=0$, in accord with the definitions of ${\cal E}, \mathbb{S}, \chi^*$.}
\begin{equation}\label{E-S-chi}
\d{}{\mathbb{S}}={-}\delta \chi^*\, , \qquad  \delta {\cal E}=\d{} \chi^*\, , \qquad \chi^*\wedge\chi^*={2\,} {\cal E}\,  {\mathbb{S}}\, ,
\end{equation}
where $\d{}$ and $\delta$ denote exterior derivative over spacetime and field space respectively. These equations specify $\d{}$ and $\delta$ of $\chi^*$ for a given ${\cal E}, {\mathbb{S}}$.  
Note that ${\cal E}_{\mu\nu}$ is the volume form of the 2-dim. space transverse to ${\cal S}$ and $\mathbb{S}$ is the volume form of the 2-dim. part of the solution space (spanned by ${\cal B}, {\cal C}$) specifying  position of the corner ${\cal S}$ on ${\cal M}$.

\section{CPSF for fluctuating boundary}\label{sec:CFSP}
In this section, we generalize CPSF within Lee-Wald \cite{Lee:1990nz} approach to cases the boundaries are permitted to fluctuate.  We extend the definition of symplectic two-form and its conservation. Within the CPSF symplectic form comes with a set of freedoms (ambiguities), in section \ref{sec:freedoms-symp-form} we discuss freedoms in CPSF with fluctuating boundaries. 

\subsection{Action and symplectic potential}\label{sec:action-symp-pot}
The action for a physical system on the manifold $\MBL$, without loss of generality, can be represented on the hypothetical boundary-less manifold $\mathcal{M}$:\footnote{$\MBL$ has a single boundary ${\cal B}$. This may be generalized to a theory on any number of boundaries, ${\cal M}_{{\cal B}_i}$, described by the  action 
$	S=\int_{{\cal M}_{{\cal B}_i}} \d{}^{D}x\, \mathcal{L}[\Phi]=\int_{\mathcal{M}} \d{}^{D}x\, \mathcal{L}[\Phi]\,  \prod_{i}^{N} H(\bdry_i)\, ,$
where $\bdry_i$ with $i=1, \cdots ,N$ denotes the boundaries of ${\cal M}_{{\cal B}_i}$.} 
\begin{equation}\label{action-original}
	S=\int_{\MBL} \d{}^{D}x\, \mathcal{L}[\Phi]=\int_{\mathcal{M}} \d{}^{D}x\, \mathcal{L}[\Phi]\,  H(\bdry)\, ,
\end{equation}
where $\mathcal{L}[\Phi]$ is Lagrangian density, $\Phi$ denotes the collection of fields and 
\begin{equation}\label{step-function}
	H(\bdry):=\left\{\begin{array}{cc} 1 & \qquad \mathcal{B}(x) \geq 0\\ 0 & \qquad \mathcal{B}(x)<0 \end{array}\right. 
\end{equation}
is the Heaviside step-function . The ``step-function  trick'' allows us to write the action on spacetime with a boundary on the spacetime ${\cal M}$ which has no boundary. 
This enables us to perform integration by-parts without worrying about total derivative terms, as $\int_{\mathcal{M}} \d{}^{D}x\, U\, \partial V=-\int_{\mathcal{M}} \d{}^{D}x\, V\, \partial U$. 

More importantly, the step-function  trick facilitates studying  variation of the action while allowing the boundary to have physical fluctuations. Explicitly,
\begin{equation}\label{delta-S-1}
	\delta S=\int_{\mathcal{M}} \d{}^{D}x\, \Big(\delta \mathcal{L}\, H(\mathcal{B})+\mathcal{L}\, \Delta(\mathcal{B})\,\delta \mathcal{B}\Big)\, ,
\end{equation}
where $\Delta(\mathcal{B})$ is the Dirac delta-function with $\delta H(\mathcal{B})= \Delta(\mathcal{B})\,\delta\mathcal{B}$.\footnote{The Dirac delta-function is usually denoted by the symbol $\delta$. However, in this context, $\delta$ is already designated as the exterior derivative on the field space. To avoid confusion, we will use the symbol $\Delta$ to denote the Dirac delta-function. Note that $\int_{\mathcal{M}}\d{}^{D}x\, \Delta(\mathcal{B})
	=\int_{\bdry}\d{}^{D-1}x$.} 
The term proportional to $H({\cal B})$ represents variation of the Lagrangian over the region $\MBL$ while the second term, which is localized on the boundary $\bdry$, appears as a result of boundary fluctuations.

Variation of the Lagrangian $\delta {\cal L}$ is given by
\begin{equation}\label{var-L}
    \delta \mathcal{L}={E}_{\Phi}[\Phi]\, \delta \Phi+\partial_{\mu}\Theta^{\mu}_{\text{\tiny{LW}}}[\delta \Phi ; \Phi]\, ,
\end{equation}
where ${E}_\Phi[\Phi]=0$ gives equations of motion.   The Lee-Wald pre-symplectic potential $\Theta^{\mu}_{\text{\tiny{LW}}}[\delta \Phi ; \Phi]$ emerges as a result of the integration by-parts of ${\cal M}$ \cite{Lee:1990nz}; it is a vector density in spacetime and a one-form in solution  space. 

Plugging \eqref{var-L} into \eqref{delta-S-1} we arrive at
\begin{equation}\begin{split}
    \delta S&=\int_{\MBL} \d{}^D x\,  E_{\Phi} \delta \Phi+\int_{\mathcal{M}} \d{}^{D}x\, \Big(\partial_{\mu}\Theta^{\mu}_{\text{\tiny{LW}}}\, H(\bdry)+\mathcal{L}\, \Delta(\bdry)\,\delta \bdry\Big)\, \\
    &=\int_{\MBL} \d{}^D x\,  E_{\Phi} \delta \Phi+\int_{\mathcal{M}} \d{}^{D}x\, \Big(-\Theta^{\mu}_{\text{\tiny{LW}}}\, \partial_{\mu}\bdry\, \Delta(\bdry)+\mathcal{L}\, \Delta(\bdry)\,\delta \bdry\Big)\, \\
    &= \int_{\MBL} \d{}^D x\,  E_{\Phi} \delta \Phi + \int_{\bdry} \d x_\mu\, \left( \Theta^{\mu}_{\text{\tiny{LW}}}[\delta \Phi ; \Phi] +N_{\bdry}\, b^\mu \, \mathcal{L} \, \delta \bdry \right)\, ,
\end{split}\end{equation}
where we have used $\partial_{\mu}H(\bdry)=\Delta(\bdry)\, \partial_{\mu}\bdry$, the definition of the normal vector on $\bdry$ \eqref{bmu-Nb} and the integral notation \eqref{notation-int}. Ultimately, by using \eqref{chi-def} we can rewrite the variation of the action as
\begin{equation}\label{delta-S}
            \tcbset{fonttitle=\scriptsize}
            \tcboxmath[colback=white,colframe=gray]{
            \delta S=\int_{\mathcal{M}} \d{}^{D}x \, E_{\Phi}\, H(\bdry)\, \delta \Phi+\int_{\bdry}\d{}x_{\mu}\, \big(\Theta^{\mu}_{\text{\tiny{LW}}}+\chi^{\mu}\, \mathcal{L}\big)\, .
            }
\end{equation}
Note that the boundary term besides the usual Lee-Wald pre-symplectic potential, has a second term which arises as a result of allowing for variations in the boundary $\delta {\cal B}$. From now on we will denote the total pre-symplectic potential as ${\Theta}^{\mu}:=\Theta^{\mu}_{\text{\tiny{LW}}}+\chi^{\mu}\, \mathcal{L}$. See \cite{Ciambelli:2021nmv, Freidel:2021dxw,Carrozza:2022xut} for a similar result.
\subsection{Symplectic form}\label{sec:symp-form}
According to the covariant phase space formalism, symplectic $(0,2)-$form of the theory \eqref{action-original} is given by variation of the integral of total pre-symplectic potential over a codimension-one hypersurface $\Bplus$ (see Fig. \ref{fig:3dsetup}) 
\begin{equation}\label{Omega-def}   \Omega=\delta\int_{\Bplus}\d{}x_{\mu}\, \big(\Theta^{\mu}_{\text{\tiny{LW}}}+\chi^{\mu}\, \mathcal{L}\big)=\delta\int_{\bdry}\d{}x_{\mu}\, \big(\Theta^{\mu}_{\text{\tiny{LW}}}+\chi^{\mu}\, \mathcal{L}\big)H(\cauchy)\, .
\end{equation}
Similarly to the previous subsection, in the second equality we employed the step-function trick to extend the integral over $\Bplus=\bdry \cap \MCL$ to the entire region $\mathcal{B}$.
\footnote{{Note that in the usual ``fixed-boundary'' CPSF, symplectic  $(0,2)-$form is defined through the integration of a symplectic current $\omega^{\mu}=\delta \Theta^{\mu}$ over a codimension-one slice. In the fluctuating boundary case, the symplectic current is not defined; instead, the consistent symplectic form should be obtained from variation of a codimension-one integral as in \eqref{Omega-def}. This guarantees the closedness of $\Omega$, $\delta\Omega=0$. For fixed boundaries, there is no difference between these methods.}}

The phase space variation $\delta$ can't be pushed through the integral onto the integrand, as it also acts on ${\cal B}$ of the integral range which exhibits a non-zero variation. To circumvent this issue, we insert a Dirac delta-function and convert the integral over $\bdry$ to an integral over ${\cal M}$,
\begin{equation}
    \Omega=\delta \int_{\mathcal{M}}\d{}^{D}x\, \big(-\partial_{\mu}\bdry\, \Theta^{\mu}_{\text{\tiny{LW}}}+\mathcal{L}\,\delta \bdry\big)H(\cauchy)\, \Delta(\bdry)\, .
\end{equation}
We can now  pass $\delta$ through the integral onto the integrand, yielding
\begin{equation}
    \begin{split}
\Omega&=\int_{\Bplus}\d{}x_{\mu}\, \delta\Theta^{\mu}_{\text{\tiny{LW}}}+ \int_{\mathcal{M}}\d{}^{D}x\, \Big[\delta \cauchy \wedge \big(-\partial_{\mu}\bdry\, \Theta^{\mu}_{\text{\tiny{LW}}}+\mathcal{L}\,\delta \bdry\big)\Delta(\bdry)\, \Delta(\cauchy)\\
        &+\big(-\partial_{\mu}\delta\bdry \wedge  \Theta^{\mu}_{\text{\tiny{LW}}}+\delta\mathcal{L}\wedge\delta \bdry\big)H(\cauchy)\, \Delta(\bdry)-\partial_{\mu}\bdry\, \delta \bdry \wedge\Theta^{\mu}_{\text{\tiny{LW}}}\, H(\cauchy)\, \Delta'(\bdry)\Big]\, .
    \end{split}
\end{equation}
Eq.\eqref{var-L} implies  $\delta \mathcal{L} =\partial_{\mu}\Theta^\mu_{\text{\tiny{LW}}}$ on-shell. Hereafter all our equalities will be on-shell. In particular, after integration by-part, the second line reduces to $\partial_{\mu} \cauchy\, \delta \bdry \wedge \Theta^{\mu}_{\text{\tiny{LW}}} \, \Delta(\cauchy) \Delta(\bdry)$ and we have
    \begin{align}
        \Omega
        &=\int_{\Bplus}\d{}x_{\mu}\, \delta\Theta^{\mu}_{\text{\tiny{LW}}}
        + \int_{\mathcal{M}}\d{}^{D}x\, \Big[ \big(-\partial_{\mu}\bdry\, \delta \cauchy \wedge\Theta^{\mu}_{\text{\tiny{LW}}}+\partial_{\mu}\cauchy\, \delta \bdry \wedge\Theta^{\mu}_{\text{\tiny{LW}}}+\mathcal{L}\, \delta \cauchy \wedge \delta \bdry\big)\Delta(\bdry)\, \Delta(\cauchy)\Big]\, \nonumber\\
&=\int_{\Bplus}\d{}x_{\mu}\, \delta\Theta^{\mu}_{\text{\tiny{LW}}}+ \int_{\cal S}\d{}^{D-2}x\, \big(-\chi^*_{\mu}\wedge\Theta^{\mu}_{\text{\tiny{LW}}}+\frac{1}{2} \chi^{\mu} \wedge \chi^*_{\mu} \,\mathcal{L} \big)\, ,
\end{align}
where in the second line we used \eqref{chi-chi*-def} and \eqref{basic-form-relation-2}, and the two Dirac delta-functions lead to a total corner term. Using \eqref{chi*-def} and  \eqref{notation-int} we arrive at
\begin{equation}\label{symp-form-var-bndy}
\tcbset{fonttitle=\scriptsize}
\tcboxmath[colback=white,colframe=gray]{
\begin{aligned} 
\Omega[\delta \Phi, \delta \Phi; \Phi]&=\Omega_{\text{\tiny{LW}}}[\delta \Phi, \delta \Phi; \Phi]+\Omega_c[\delta \Phi, \delta \Phi; \Phi]\, ,\\
       \Omega_{\text{\tiny{LW}}}[\delta \Phi, \delta \Phi; \Phi] &:=\int_{\bdry}\d{}x_{\mu}\, \delta\Theta^\mu_{\text{\tiny{LW}}}\  H(\cauchy)\, ,  \\
       \Omega_c[\delta \Phi, \delta \Phi; \Phi] &:=
       \int_{\cal S}\d{}x_{\mu\nu}\, \big(2\, \Theta^{\mu}_{\text{\tiny{LW}}}\wedge \chi^{\nu}+\mathcal{L}\, \chi^{\mu}\wedge \chi^{\nu}\big)\, ,            
\end{aligned}
}
\end{equation}
where $\Omega_{\text{\tiny{LW}}}$  is the standard Lee-Wald symplectic two-form \cite{Lee:1990nz} and $\Omega_c$ which involves $\delta{\cal A}, \delta \bdry, \delta\cauchy$ terms arises from allowing the boundaries to vary. As the name suggests, $\Omega_c$ is a codimension-two corner integral. In the usual terminology of \cite{Adami:2021kvx, Sheikh-Jabbari:2022mqi, Adami:2023wbe}, $\Omega_{\text{\tiny{LW}}}$ involves bulk modes (like gravitational waves) as well as boundary modes, whereas $\Omega_c$ only involves boundary modes. Thus, as expected, this confirms the expectation that ${\cal A},  \bdry, \cauchy$ are boundary modes. 

\subsection{Conservation of symplectic form}\label{sec:conservation-symp-form}
\begin{figure}
    \centering
    \includegraphics[scale=.6]{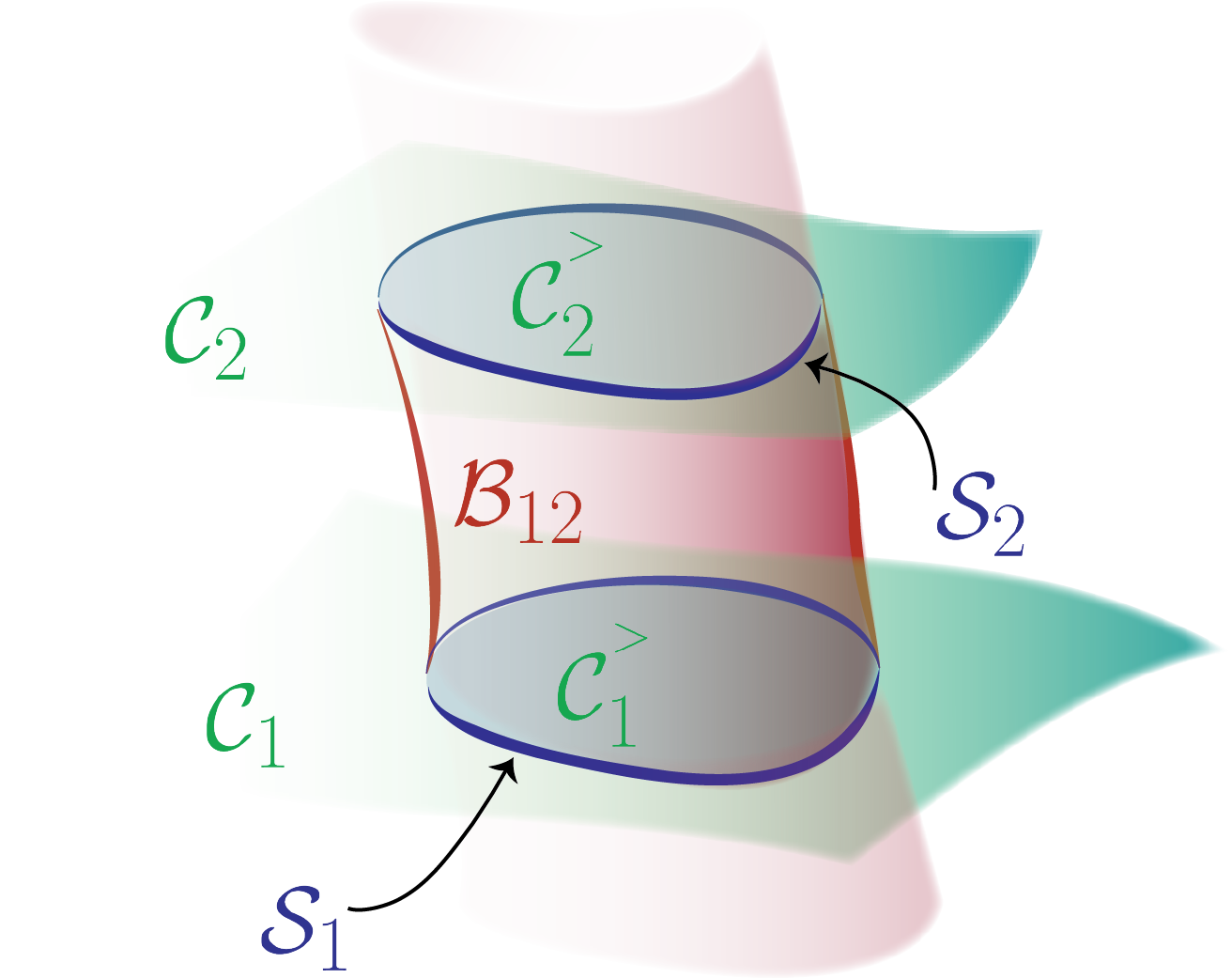}
    \caption{$\textcolor{Green}{\cauchy_1}$ and $\textcolor{Green}{\cauchy_2}$ are two (partial) Cauchy surfaces and $\textcolor{BrickRed}{{\cal B}_{12}}$ is the causal boundary of the region of spacetime bounded between $\textcolor{Green}{\cauchy_1}$ and $\textcolor{Green}{\cauchy_2}$. The intersection of $\textcolor{BrickRed}{\bdry_{12}}$, $\textcolor{Green}{\cauchy_1}$, and $\textcolor{Green}{\cauchy_2}$  creates two codimension-two corners $\textcolor{Blue}{\mathcal{S}_1}$ and $\textcolor{Blue}{\mathcal{S}_2}$ indicated by dark blue lines.
    $\textcolor{Green} {\cauchy_1^{^>}}$ and $\textcolor{Green}{\cauchy_2^{^>}}$ represent the portions of $\textcolor{Green}{\cauchy_1}$ and $\textcolor{Green}{\cauchy_2}$ that are enclosed by these corners.
    }
    \label{fig:3dconservation}
\end{figure}

In this subsection, we study the conservation of the symplectic two-form. To do so, we start with the boundary-less manifold $\mathcal{M}$ now with three embedding codimension-one hypersurfaces $\bdry$, $\cauchy_1$, and $\cauchy_2$ (see Fig.~\ref{fig:3dconservation}). The action in the region enclosed by these hypersurfaces is given as follows
\begin{equation}
    S=\int_{\mathcal{M}} \d{}^{D}x\, \mathcal{L}\, H(\bdry)\, H(\cauchy_1)\, H(-\cauchy_2)\, .
\end{equation}
Using \eqref{var-L} the first variation of the action is given by
\begin{equation}
    \begin{split}
        \delta S 
        &=\int_{\mathcal{M}} \d{}^{D}x\, E_{\Phi}\, \delta \Phi H(\bdry)\, H(\cauchy_1)\, H(-\cauchy_2)\\
        &+\int_{\bdry}\d{}^{D-1}x\, \Theta_{\bdry}\, H(\cauchy_1)\, H(-\cauchy_2)+\int_{\cauchy_1}\d{}^{D-1}x\, \Theta_{\cauchy_1}\, H(\bdry) -\int_{\cauchy_2}\d{}^{D-1}x\, \Theta_{\cauchy_2}\, H(\bdry)\, \, ,
    \end{split}
\end{equation}
where
\begin{equation}
    \begin{split}
        &\Theta_{\bdry}:=-\partial_{\mu}\bdry\, \Theta^{\mu}_{\text{\tiny{LW}}}+\mathcal{L}\, \delta \bdry\, ,\qquad
        \Theta_{\cauchy_i}:=-\partial_{\mu}\cauchy_i\, \Theta^{\mu}_{\text{\tiny{LW}}}+\mathcal{L}\, \delta \cauchy_i\, 
        , \quad i=1,2\, ,
    \end{split}
\end{equation}
and in the last line we have used $H(\cauchy_1)\big|_{\cauchy_2}=1$ and $H(-\cauchy_2)\big|_{\cauchy_1}=1$. 
Using $\delta^{2}S=0$ and assuming that $\Phi$ and $\delta\Phi$ are respectively satisfied equations of motion, $E_{\Phi}=0$, and linearized equations of motion, $\delta E_{\Phi}=0$, we get
\begin{equation}
    \delta\int_{\bdry}\d{}^{D-1}x\, \Theta_{\bdry}\, H(\cauchy_1)\, H(-\cauchy_2)+\delta\int_{\cauchy_1}\d{}^{D-1}x\, \Theta_{\cauchy_1}\, H(\bdry)-\delta\int_{\cauchy_2}\d{}^{D-1}x\, \Theta_{\cauchy_2}\, H(\bdry)=0\, .
\end{equation}
 One can rewrite this equation as follows
\begin{equation}\label{symp-form-cons}
    \Omega(\cauchy_2^{^>})-\Omega(\cauchy_1^{^>})=\Omega(\bdry_{12})\, ,
\end{equation}
where 
\begin{equation}
    \begin{split}
\Omega(\bdry_{12}):=\delta\int_{\bdry}\d{}^{D-1}x\, \Theta_{\bdry}\, H(\cauchy_1)\, H(-\cauchy_2)\, ,\qquad \Omega(\cauchy_i^{^>}):=\delta\int_{\cauchy_i}\d{}^{D-1}x\, \Theta_{\cauchy_i}\, H(\bdry), \quad i=1,2\, .
    \end{split}
\end{equation}
Equation \eqref{symp-form-cons} represents the conservation of the symplectic two-form. In our previous subsections, we specifically focused on $\Omega(\bdry_{12})$. More precisely, $\Omega(\bdry_{12})$ exhibits two corners ($\mathcal{S}_1$ and $\mathcal{S}_2$) resulting from the intersection of $\bdry$ with $\cauchy_1$ and $\cauchy_2$ (as depicted in Fig.~\ref{fig:3dconservation}). 

\subsection{Freedoms in symplectic form}\label{sec:freedoms-symp-form}
In CPSF, pre-symplectic potential and symplectic form are specified up to two class of freedoms (usually called ``ambiguities''), see \cite{Iyer:1994ys, Grumiller:2022qhx}. 

\paragraph{$W$-freedom.} It is well known that Lagrangians differing up to total derivative terms  $\mathcal{L} \rightarrow \mathcal{L} + \partial_\mu W^\mu$ yield the same equations of motion and are hence classically equivalent. 

\paragraph{$Y$-freedom.} The definition of pre-symplectic potential \eqref{var-L} and that $\delta {\cal L}=\partial_\mu\Theta^\mu$ on-shell, determines $\Theta^\mu$ up to a total divergence term,  $\Theta^{\mu}_{\text{\tiny{LW}}} \rightarrow \Theta^{\mu}_{\text{\tiny{LW}}}+\partial_{\nu}Y^{\mu\nu}$, where $Y^{\mu\nu}$ is a skew-symmetric tensor and a one-form over the field space, and is constructed out of fields $\Phi$ and their variations $\delta \Phi$. 

By putting all freedoms together we find the following general form of the pre-symplectic potential
\begin{equation}\label{symp-pot-freedom}
            \tcbset{fonttitle=\scriptsize}
            \tcboxmath[colback=white,colframe=gray]{
            \tilde{\Theta}^{\mu}=\Theta^{\mu}+ \delta W^\mu + \chi^\mu\, \partial_\nu W^\nu+\partial_{\nu}Y^{\mu\nu}\, .
            }
\end{equation}
Having the general form of the symplectic potential \eqref{symp-pot-freedom}, we can compute the corresponding symplectic two-form
\begin{equation}
    \tilde{\Omega}=\Omega+\delta\int_{{\cal B}}\d{}x_{\mu}\left(\delta W^\mu + \chi^\mu\, \partial_\nu W^\nu+\partial_{\nu}Y^{\mu\nu}\right)H({\cal C})\, .
\end{equation}
We use identities in appendix \ref{Appen:variations}, in particular \eqref{codim1-1form}, to evaluate the freedoms in the symplectic two-form. We start with the $W$ terms
\begin{equation}
    \begin{split}
        \delta\int_{\bdry} \d{}x_{\mu}\, (\delta W^\mu + \chi^\mu\, \partial_\nu W^\nu) \, H(\cauchy)&=\int_{\bdry} \d{}x_{\mu}\, \qty[\delta \chi^{\mu} - \partial_{\nu} \chi^{\mu} \wedge \chi^{\nu} ]\partial_\alpha W^\alpha H(\cauchy)\\
        \hspace{5mm}&+\int_{{\cal S}}\d{}x_{\mu\nu}\, \qty[2\delta W^{\mu}  \wedge \chi^{\nu}+{\chi^\mu \wedge \chi^\nu \partial_\alpha W^\alpha}]\, .
    \end{split}
\end{equation}
The first integral on the RHS vanishes, cf. \eqref{b-del-chi}. 
Next, we add the $ Y$ piece {(using  integration by-parts and \eqref{identity-delta-2-1})}, 
\begin{equation}
            \tilde{\Omega}=\Omega+\int_{{\cal S}}\d{}x_{\mu\nu}\left[2\delta W^{\mu}\wedge \chi^{\nu}-\delta Y^{\mu\nu} +(2\partial_{\alpha}Y^{\mu\alpha}+\partial_{\alpha}W^{\alpha}\, \chi^{\mu})\wedge \chi^{\nu}\right]\, .
\end{equation}
Straightforward, but lengthy algebra yields (details in \eqref{Omega-freedom}), 
\begin{equation}\label{symp-form-freedom}
            \tcbset{fonttitle=\scriptsize}
            \tcboxmath[colback=white,colframe=gray]{
            \tilde{\Omega}[\delta \Phi, \delta \Phi; \Phi]=\Omega-\int_{{\cal S}}\d{}x_{\mu\nu}\left(\delta \bar{Y}^{\mu\nu} -2\partial_{\alpha}\bar{Y}^{\mu\alpha}\wedge \chi^{\nu}\right)=\Omega[\delta \Phi, \delta \Phi; \Phi]-\delta \int_{{\cal S}}\d{}x_{\mu\nu}\, \bar{Y}^{\mu\nu} \, ,
            }
\end{equation}
where 
\begin{equation}
    \bar{Y}^{\mu\nu}:=Y^{\mu\nu}-2 W^{[\mu}\ \chi^{\nu]}\, ,
\end{equation}
and in the second equality, we used identity \eqref{identity-delta-2-1}. 
That is, only the combination $\bar{Y}$ appears in the shift of the symplectic form and for any $W$-freedom one can choose an appropriate $Y$ term, such that the contribution of $W$ in the symplectic form drops out. This result dovetails with the usual statement that in the absence of boundary variations, $W$-freedom does not change the symplectic form,\footnote{Note that in the absence of boundary fluctuations $\chi=0$ and hence $\bar{Y}=Y$.} while $Y$-freedom shifts $\Omega[\delta \Phi, \delta \Phi; \Phi]$.

\section{Surface charge analysis }\label{sec:surface-charge}
In this section, we compute surface charges associated with local symmetries in CPSF with fluctuating boundaries. In particular, we show that surface charges are integrable and moreover, show that the integrable charge is the Noether charge. We also analyze the charge algebra and compute central extension terms. We discuss conservation of the charges and what they mean in the fluctuating boundary case.

\subsection{Surface charges in fluctuating boundary case are integrable}\label{sec:charges-are-integrable}

Given the symplectic two-form \eqref{symp-form-var-bndy} one can associate a surface charge  variation $\slashed{\delta}Q(\xi)$ to  diffeomorphisms $\xi^\mu$, explicitly,  $\slashed{\delta}Q(\xi)=\Omega[\delta_\xi \Phi, \delta \Phi; \Phi]$.  $\slashed{\delta}Q(\xi)$ is a one-form over the field space and in general is not an exact one-form. If it is exact then the charge is integrable. In what follows we establish that the charges are \textit{always} integrable. 

Starting from \eqref{symp-form-var-bndy}, we decompose the surface charge variation into the Lee-Wald and corner contributions and  compute them separately, 
\begin{equation}\label{total-Q}
    \slashed{\delta}Q(\xi)=\slashed{\delta}Q_{\text{\tiny{LW}}}(\xi)+ \slashed{\delta}Q_{c}(\xi)\, ,
\end{equation}
where
\begin{equation}\label{Charge-variation-1}
        \slashed{\delta}Q_{\text{\tiny{LW}}}(\xi):=\Omega_{\text{\tiny{LW}}}[\delta_\xi \Phi, \delta \Phi; \Phi]\, ,\qquad  
        \slashed{\delta}Q_{c}(\xi):=\Omega_c[\delta_{\xi} \Phi, \delta \Phi; \Phi]\, . 
\end{equation}
\paragraph{Lee-Wald part.} Let us focus on the Lee-Wald  contribution in \eqref{Charge-variation-1}, in the terminology of \cite{Lee:1990nz},
\begin{equation}\label{Q-LW}
    \slashed{\delta}Q_{\text{\tiny{LW}}}(\xi)=-\int_{{\cal S}}\d{}x_{\mu\nu}\big(\delta Q_{\text{\tiny{N}}}^{\mu\nu}(\xi)+2\, \xi^{\mu}\, \Theta^{\nu}_{\text{\tiny{LW}}}[\delta\Phi;\Phi]\big)\, .
\end{equation}
Here we assumed $\xi$ are field independent, one can simply generalize the above argument for field dependent $\xi$ and $Q_{\text{\tiny{N}}}^{\mu\nu}(\xi)$ is the Noether charge density and is related to the Noether current $J^\mu_{\xi}$,
\begin{equation}\label{Noether-current}
    J^\mu_{\xi}:=\Theta^{\mu}_{\text{\tiny{LW}}}[\delta_{\xi} \Phi; \Phi]-\xi^{\mu}\, \mathcal{L}\, , \qquad J^\mu_{\xi}=\partial_{\nu}Q_{\text{\tiny{N}}}^{\mu\nu}(\xi)\, .
\end{equation}
The second term in \eqref{Q-LW} is  the source of non-integrability in the Lee-Wald surface charge variation, in the fixed-boundary cases where  we assume ${\cal S}$ and $ \d{}x_{\mu\nu}$ have zero variations over the field space. In our setting, however, we allow them to have nonzero variations and as we will show, this renders the charge integrable.

\paragraph{Corner part.} The corner term in  \eqref{Charge-variation-1} yields
\begin{equation}\label{Q-xi-c-1}
    \slashed{\delta}Q_{c}(\xi)=\int_{{\cal S}} \d{}x_{\mu\nu}\, \Big[-2(\mathcal{L}\, \chi^{\mu}[\delta_\xi \Phi; \Phi]+\Theta^{\mu}_{\text{\tiny{LW}}}[\delta_\xi \Phi; \Phi])\, \chi^{\nu}[\delta \Phi; \Phi]+2\,\Theta^{\mu}_{\text{\tiny{LW}}}[\delta \Phi; \Phi]\, \chi^{\nu}[\delta_\xi \Phi;\Phi]\Big]\, .
\end{equation}
We compute $\chi^{\mu}[\delta_\xi \Phi; \Phi]$ using  \eqref{chi-def}, 
\begin{equation}
\begin{split}
 \chi^{\mu}[\delta_\xi \Phi; \Phi]&=-N\, {n^{\mu}}\, \delta_{\xi} \cauchy + N_{\bdry}\, {b^{\mu}}\, \delta_{\xi} \bdry+ N\, \mathcal{A}\, n^{\mu}\, \delta_{\xi} \bdry\, ,\\
    &=-( b^{\mu}\, b_{\nu}- n^{\mu}\, n_{\nu})\xi^{\nu}=-\xi^{\mu}+q^{\mu}_{\nu}\xi^{\nu}\, , 
\end{split}
\end{equation}
where we used
\begin{equation}
        \delta_{\xi} \bdry=\mathcal{L}_{\xi}\bdry=\xi^{\mu}\partial_{\mu}\bdry=-N_{\bdry}^{-1}\, b_{\mu}\, \xi^{\mu}\, ,\qquad 
        \delta_{\xi} \cauchy=\mathcal{L}_{\xi}\cauchy=\xi^{\mu}\partial_{\mu}\cauchy=-N_{\cauchy}^{-1}\, c_{\mu}\, \xi^{\mu}\, .
\end{equation}
Thus, by using the definition of the Noether current \eqref{Noether-current} and  $\d{}x_{\mu\nu}\, q^{\mu}_{\alpha}\, \xi^{\alpha}=0$,
we have
\begin{align}\label{Qc}
\slashed{\delta}Q_{c}(\xi)
&=-\int_{{\cal S}} \d{}x_{\mu\nu}\, \left(2\, J^{\mu}_{\xi}\, \chi^{\nu}[\delta\Phi;\Phi]+2\, \Theta^{\mu}_{\text{\tiny{LW}}}[\delta\Phi;\Phi]\, \xi^{\nu}\right)\,.
\end{align}

Putting  \eqref{Q-LW} and \eqref{Qc} together, we obtain
\begin{equation}\label{total-charge}
           {
            \slashed{\delta}Q(\xi)=-\int_{{\cal S}} \d{}x_{\mu\nu}\, (\delta Q_{\text{\tiny{N}}}^{\mu\nu}(\xi)+2J^{\mu}_{\xi}\, \chi^{\nu})\, .
            }
\end{equation}
Using  \eqref{bi-normal},  \eqref{basic-form-relation-1} and \eqref{basic-form-relation-2} we arrive at
\begin{equation}\label{Charge-variation-2}
    \begin{split}
        \slashed{\delta}Q(\xi)&=-\frac{1}{2}\int_{{\cal S}} \d{}^{D-2}x\,  \left[\delta ({\cal E}_{\mu\nu}\, Q_{\text{\tiny{N}}}^{\mu\nu})- Q_{\text{\tiny{N}}}^{\mu\nu}\, \delta {\cal E}_{\mu\nu} +2 J^{\mu}_{\xi}\, \chi^*_{\mu} \right]\, \\
        &=-\frac{1}{2}\int_{{\cal S}} \d{}^{D-2}x\,  \left[\delta ({\cal E}_{\mu\nu}\, Q_{\text{\tiny{N}}}^{\mu\nu}) + 2\partial_{\nu}(Q_{\text{\tiny{N}}}^{\mu\nu}\, \chi^*_{\mu})\right]\, \\
        & =-\frac{1}{2}\int_{{\cal S}} \d{}^{D-2}x\,  \left[\delta ({\cal E}_{\mu\nu}\, Q_{\text{\tiny{N}}}^{\mu\nu}) + \partial_{\alpha}({\cal E}_{\mu\nu}\, Q_{\text{\tiny{N}}}^{\mu\nu}\, \chi^{\alpha})-2\partial_{\alpha}( Q_{\text{\tiny{N}}}^{\mu\nu}\, q^\alpha_{\mu} \chi^*_{\nu})\right]\, ,
    \end{split}
\end{equation}
where in the last line we used the identity \eqref{identity-E-chi}, ${\cal E}_{\mu\nu}\, \chi^{\alpha}=2 q^\alpha_{[\mu} \chi^*_{\nu]} - 2 \delta^\alpha_{[\mu} \chi^*_{\nu]}$, which yields,
$$    Q_{\text{\tiny{N}}}^{\mu\nu}\, {\cal E}_{\mu\nu}\, \chi^{\alpha}=2 Q_{\text{\tiny{N}}}^{\mu\nu}\, q^\alpha_{\mu} \chi^*_{\nu} + 2 Q_{\text{\tiny{N}}}^{\mu\nu}\, \delta^\alpha_{\nu} \chi^*_{\mu}\, .
$$
The last term in the third line of \eqref{Charge-variation-2} is a total derivative term over $\cal S$ and assuming $\cal S$ is a compact surface, we drop it. Employing identity \eqref{identity-2}, we find 
\begin{equation}
 {\delta}Q(\xi)= - \, \delta\int_{\cal S}\d{}x_{\mu\nu}\, Q_{\text{\tiny{N}}}^{\mu\nu}(\xi)\, .
\end{equation}
The above is the main result of this subsection. It is in accord with the results of \cite{Ciambelli:2021nmv, Freidel:2021dxw, Carrozza:2022xut} and establishes that the surface charge expressions are \textit{integrable}. Finally, by integrating over the phase space, we get
\begin{equation}\label{surface-charge-integrable}
            \tcbset{fonttitle=\scriptsize}
            \tcboxmath[colback=white,colframe=gray]{
            Q(\xi)= - Q_{\text{\tiny{N}}}(\xi)+Q_{_{\cal R}}(\xi)\, , \qquad   Q_{\text{\tiny{N}}}(\xi)=:  \int_{{\cal S}}\d{}x_{\mu\nu}\, Q_{\text{\tiny{N}}}^{\mu\nu}(\xi)\, .
            }
\end{equation}
where $Q_{_{\cal R}}(\xi)$ is a constant over the field space, $\delta Q_{_{\cal R}}(\xi)=0$.  $Q_{_{\cal R}}(\xi)$  is linear in $\xi$ and may be viewed as the charge at a reference point. {In other words, if we denote the zero-charge reference point by ${\Phi}_{_{\cal R}}$,  $Q[{\Phi}_{_{\cal R}};\xi]=0$, fixes $Q_{_{\cal R}}(\xi)=Q_{\text{\tiny{N}}}({\Phi}_{_{\cal R}};\xi)$. 

\paragraph{Freedom in defining surface charges.} As discussed in section \ref{sec:freedoms-symp-form}, we have a freedom in defining symplectic form up to an antisymmetric bivector $\bar{Y}^{\mu\nu}$ \eqref{symp-form-freedom}.  The surface charge expression \eqref{surface-charge-integrable} is also affected by the freedoms, but in a way that it remains integrable. Explicitly,  starting from \eqref{symp-form-freedom} and the definition of the Noether charge \eqref{Noether-current}, we obtain
\begin{equation}\label{charge-freedom}
            \tcbset{fonttitle=\scriptsize}
            \tcboxmath[colback=white,colframe=gray]{
           \tilde{Q}(\xi)=-\int_{{\cal S}}\d{}x_{\mu\nu}\, \left(Q_{\text{\tiny{N}}}^{\mu\nu}(\xi)+ \bar{Y}^{\mu\nu}[\delta_{\xi}\Phi;\Phi]\right) + \tilde{Q}_{_{\cal R}}(\xi)={Q}(\xi)-\int_{{\cal S}}\d{}x_{\mu\nu}\, \bar{Y}^{\mu\nu}[\delta_{\xi}\Phi;\Phi]\, .
            }
\end{equation}

\subsection{Charge algebra}\label{sec:charge-algebra}

Having the charges, we will now work out  the charge algebra recalling its definition, 
\begin{equation}
    \begin{split}
        \big{\{}Q(\xi_1),Q(\xi_2)\big{\}}:=\delta_{\xi_2}Q(\xi_1)&=- \delta_{\xi_2}\int_{{\cal S}}\d{}x_{\mu\nu}\, Q_{\text{\tiny{N}}}^{\mu\nu}(\xi_1)\, .
    \end{split}
\end{equation}
To this end, we start with the third line of the surface charge variation \eqref{Charge-variation-2},
\begin{equation}
    \delta Q(\xi)=-\frac{1}{2}\int_{{\cal S}} \d{}^{D-2}x\, \Big(\delta\mathcal{Q}_{\text{\tiny{N}}}(\xi)+\partial_{\mu}\big(\mathcal{Q}_{\text{\tiny{N}}}(\xi)\, \chi^{\mu}\big)\Big)\, , \qquad \mathcal{Q}_{\text{\tiny{N}}}(\xi):= \mathcal{E}_{\mu\nu}\, Q^{\mu\nu}_{\text{\tiny{N}}}(\xi)\, ,
\end{equation}
and compute $\delta_{\xi_2}Q(\xi_1)$,
\begin{equation}
    \begin{split}
        \delta_{\xi_2}Q(\xi_1)
        &=-\frac{1}{2}\int_{{\cal S}} \d{}^{D-2}x\, \Big(\delta_{\xi_2}\mathcal{Q}_{\text{\tiny{N}}}(\xi_1)+\partial_{\mu}\big(\mathcal{Q}_{\text{\tiny{N}}}(\xi_1)\, \chi^{\mu}[\delta_{\xi_2}\Phi;\Phi]\big)\Big)\\
        &=-\frac{1}{2}\int_{\cal S} \d{}^{D-2}x\, \Big(\mathcal{L}_{\xi_2}\mathcal{Q}_{\text{\tiny{N}}}(\xi_1)-\mathcal{Q}_{\text{\tiny{N}}}(\mathcal{L}_{\xi_2}\xi_1)-\partial_{\mu}\big(\mathcal{Q}_{\text{\tiny{N}}}(\xi_1)\, (\xi_2^{\mu}-q^{\mu}_{\alpha}\, \xi_2^{\alpha})\big)\Big)\\
        &=-\frac{1}{2}\int_{{\cal S}} \d{}^{D-2}x\, \Big(\partial_{\mu}\big(\xi^{\mu}_2\mathcal{Q}_{\text{\tiny{N}}}(\xi_1)\big)-\mathcal{Q}_{\text{\tiny{N}}}(\textrm{[}\xi_2,\xi_1\textrm{]})-\partial_{\mu}\big(\mathcal{Q}_{\text{\tiny{N}}}(\xi_1)\, (\xi_2^{\mu}-q^{\mu}_{\alpha}\, \xi_2^{\alpha})\big)\Big)\\
        &=\frac{1}{2}\int_{{\cal S}} \d{}^{D-2}x\, \mathcal{Q}_{\text{\tiny{N}}}(\textrm{[}\xi_2,\xi_1 \textrm{]})=\int_{{\cal S}}\d{}x_{\mu\nu}\,  Q^{\mu\nu}_{\text{\tiny{N}}}(\textrm{[}\xi_2,\xi_1\textrm{]})\\
        &=-Q(\textrm{[}\xi_2,\xi_1\textrm{]})+Q_{_{\mathcal{R}}}(\textrm{[}\xi_2,\xi_1\textrm{]})\, ,
    \end{split}
\end{equation}
where $\textrm{[}\xi_2,\xi_1\textrm{]}:=\mathcal{L}_{\xi_2}\xi_1$ is the Lie bracket.
In the second line, we  used $\delta \xi=0$, as also mentioned in \cite{Wald:1999wa}
\be\label{Noether-charge-covariant}
\delta_{\xi_2}\mathcal{Q}_{\text{\tiny{N}}}(\xi_1)=\mathcal{L}_{\xi_{2}} \mathcal{Q}_{\text{\tiny{N}}}(\xi_1)
-\mathcal{Q}_{\text{\tiny{N}}}\qty(\mathcal{L}_{\xi_{2}}\xi_1)\, ,
\ee  
that $\mathcal{Q}_{\text{\tiny{N}}}(\xi)$ is linear w.r.t $\xi$, and in the last line, we have dropped a total divergence term over ${\cal S}$ upon the assumption that ${\cal S}$ is compact and integrand is smooth over ${\cal S}$. Therefore, the charge algebra becomes
\begin{equation}\label{charge-algebra}
            \tcbset{fonttitle=\scriptsize}
            \tcboxmath[colback=white,colframe=gray]{
            \big{\{}Q(\xi_1),Q(\xi_2)\big{\}}=Q(\textrm{[}\xi_1,\xi_2\textrm{]})-Q_{_{\mathcal{R}}}(\textrm{[}\xi_1,\xi_2\textrm{]})\, .
            }
\end{equation}
As we will show in the next subsection, the last term yields the central extension term. The charge algebra is the same as the algebra of symmetry generators up to central extension terms (in accord with the fundamental theorem of CPSF \cite{Grumiller:2022qhx, Compere:2019qed, Seraj:2016cym}).

We close this part by the comment that, even though the charges are affected by the $\bar{Y}$-freedom (cf. \eqref{charge-freedom}), if we have ``covariant $\bar{Y}$-freedom'', i.e.  for a $\xi$-independent $\bar{Y}^{\mu \nu}$, $\delta_{\xi}\bar{Y}^{\mu\nu}=\mathcal{L}_{\xi}\bar{Y}^{\mu\nu}$,  the charge algebra is not affected by $\bar{Y}$-freedom.
\subsection{Central charge}\label{sec:central-charge} 
In this subsection, we make it explicit that the last term $Q_{_{\mathcal{R}}}(\textrm{[}\xi_1,\xi_2\textrm{]})$ in the charge algebra \eqref{charge-algebra} gives rise to the central charge. The key point in the analysis below is in the comments below \eqref{surface-charge-integrable}: Although $Q_{_{\mathcal{R}}}(\xi)$ is field-independent, it depends on the reference point field ${\Phi}_{_{\cal R}}$. Therefore, while $\delta Q_{_{\mathcal{R}}}(\xi)=0$, in general ${\delta}_{\xi_2} Q_{_{\mathcal{R}}}(\xi_1)\neq 0$. Let us call any quantity $X(\xi)$  covariant if for any $\delta \xi=0$, 
\be \label{def-covariant}
\delta_{\xi_{2}} X(\xi_1):=\delta\qty(X(\xi_1))\mid_{_{\xi_{2}}}= \mathcal{L}_{\xi_{2}} X(\xi_1)-X\qty(\mathcal{L}_{\xi_{2}}\xi_1)\, .
\ee
If $\delta_{\xi_{2}} X(\xi_1) \neq \mathcal{L}_{\xi_{2}}X (\xi_1)-X\qty(\mathcal{L}_{\xi_{2}}\xi_1)$, we say there is an anomaly (in diffeomorphism $\xi$), see \cite{Chandrasekaran:2020wwn} for a related discussion. With this definition and as \eqref{Noether-charge-covariant} shows  
Noether charge $\mathcal{Q}_{\text{\tiny{N}}}(\xi)$ is covariant. 
Central charges are typically arising from such anomalies.

As the first consequence of the above observation and discussions, we note that, since $Q_{\text{\tiny{N}}}$ is by definition covariant \eqref{Noether-charge-covariant}, then $Q(\xi)$ does not transform covariantly in the sense mentioned above. To this end, we note that
\begin{equation}
    \delta_{\xi}Q(\zeta)=-\delta_{\xi}Q_{\text{\tiny{N}}}(\zeta)=-\mathcal{L}_{\xi}Q_{\text{\tiny{N}}}(\zeta) +Q_{\text{\tiny{N}}}(\mathcal{L}_{\xi}\zeta)= \mathcal{L}_{\xi}Q(\zeta)-Q(\mathcal{L}_{\xi}\zeta)-\mathcal{L}_{\xi}Q_{_{\mathcal{R}}}(\zeta)+Q_{_{\mathcal{R}}}(\mathcal{L}_{\xi}\zeta)\, .
\end{equation}
In other words, the choice of reference points can break spacetime covariance.
Let us now examine the charge algebra \eqref{charge-algebra}. The non-covariance in the LHS is coming from both the charge $Q(\xi)$ as well as the bracket itself. In the RHS we have $Q_{_{\text{N}}}$, which is covariant. The main objective here is to extract out the non-covariance coming from the bracket, which is generically called ``central charge''. To this end, we rewrite the LHS of \eqref{charge-algebra} in terms of ``covariant part'' $Q_{\text{\tiny{N}}}(\xi)$:
\begin{equation}
    \big{\{}Q(\xi_1),Q(\xi_2)\big{\}}=\big{\{}Q_{\text{\tiny{N}}}(\xi_1),Q_{\text{\tiny{N}}}(\xi_2)\big{\}}-\big{\{}Q_{\text{\tiny{N}}}(\xi_1),Q_{_{\mathcal{R}}}(\xi_2)\big{\}}-\big{\{}Q_{_{\mathcal{R}}}(\xi_1),Q_{\text{\tiny{N}}}(\xi_2)\big{\}}+\big{\{}Q_{_{\mathcal{R}}}(\xi_1),Q_{_{\mathcal{R}}}(\xi_2)\big{\}}\, .
\end{equation}
Now we compute each term on the RHS of the above equation
\begin{equation}
    \begin{split}
        -\big{\{}Q_{\text{\tiny{N}}}(\xi_1),Q_{_{\mathcal{R}}}(\xi_2)\big{\}}=\big{\{}Q(\xi_1)-Q_{_{\mathcal{R}}}(\xi_1),Q_{_{\mathcal{R}}}(\xi_2)\big{\}} & =-\delta_{\xi_1}Q_{_{\mathcal{R}}}(\xi_2)-\big{\{}Q_{_{\mathcal{R}}}(\xi_1),Q_{_{\mathcal{R}}}(\xi_2)\big{\}}\\
        &=-\big{\{}Q_{_{\mathcal{R}}}(\xi_1),Q_{_{\mathcal{R}}}(\xi_2)\big{\}}\, ,
    \end{split}
\end{equation}
where we have used $\delta_{\xi_1}Q_{_{\mathcal{R}}}(\xi_2)=0$. The same calculation yields
\begin{equation}
    -\big{\{}Q_{_{\mathcal{R}}}(\xi_1),Q_{\text{\tiny{N}}}(\xi_2)\big{\}}=-\big{\{}Q_{_{\mathcal{R}}}(\xi_1),Q_{_{\mathcal{R}}}(\xi_2)\big{\}}\, .
\end{equation}
Putting all of these together, we find
\begin{equation}
    \big{\{}Q(\xi_1),Q(\xi_2)\big{\}}=\big{\{}Q_{\text{\tiny{N}}}(\xi_1),Q_{\text{\tiny{N}}}(\xi_2)\big{\}}-\big{\{}Q_{_{\mathcal{R}}}(\xi_1),Q_{_{\mathcal{R}}}(\xi_2)\big{\}}\, .
\end{equation}
Hence the algebra of covariant charge $Q_{\text{\tiny{N}}}$  becomes
\begin{equation}
    \big{\{}Q_{\text{\tiny{N}}}(\xi_1),Q_{\text{\tiny{N}}}(\xi_2)\big{\}}=Q_{\text{\tiny{N}}}(\textrm{[}\xi_1,\xi_2\textrm{]})+\big{\{}Q_{_{\mathcal{R}}}(\xi_1),Q_{_{\mathcal{R}}}(\xi_2)\big{\}}\, .
\end{equation}
We compute the last term  by evaluating the above equation at the reference point ${\cal R}$
\begin{equation}
    \big{\{}Q_{\text{\tiny{N}}}(\xi_1),Q_{\text{\tiny{N}}}(\xi_2)\big{\}}\big|_{_{\mathcal{R}}}=Q_{_{\mathcal{R}}}(\textrm{[}\xi_1,\xi_2\textrm{]})+\big{\{}Q_{_{\mathcal{R}}}(\xi_1),Q_{_{\mathcal{R}}}(\xi_2)\big{\}}\, ,
\end{equation}
where we used $Q_{\text{\tiny{N}}}\big|_{\mathcal{R}}=Q_{_{\mathcal{R}}}$. 
Finally, the charge algebra becomes \footnote{Eq.~\eqref{charge-algebra-N} and that $\mathcal{K}[\xi_1;\xi_2]$ is a central charge, may also be written as
$    \int_{\gamma} \delta\Big(\big{\{}Q_{\text{\tiny{N}}}(\xi_1),Q_{\text{\tiny{N}}}(\xi_2)\big{\}}-Q_{\text{\tiny{N}}}(\textrm{[}\xi_1,\xi_2\textrm{]})\Big)=0, 
$ 
where $\int_{\gamma}$ is an integral over path $\gamma$ in the phase space which connects $\Phi$ to the reference point ${\Phi}_{_{\cal R}}$.}
\begin{equation}\label{charge-algebra-N}
            \tcbset{fonttitle=\scriptsize}
            \tcboxmath[colback=white,colframe=gray]{
              \begin{aligned}
        &\big{\{}Q_{\text{\tiny{N}}}(\xi_1),Q_{\text{\tiny{N}}}(\xi_2)\big{\}}=Q_{\text{\tiny{N}}}(\textrm{[}\xi_1,\xi_2\textrm{]})+\mathcal{K}[\xi_1;\xi_2]\, ,\\
        & \mathcal{K}[\xi_1;\xi_2]:=\big{\{}Q_{\text{\tiny{N}}}(\xi_1),Q_{\text{\tiny{N}}}(\xi_2)\big{\}}\big|_{_{\mathcal{R}}}-Q_{_{\mathcal{R}}}(\textrm{[}\xi_1,\xi_2\textrm{]})\, .
    \end{aligned}
            }
\end{equation}
The central extension term, $\mathcal{K}[\xi_1;\xi_2]$, emerges explicitly in the derivation due to non-covariance (see \cite{Chandrasekaran:2020wwn, Rignon-Bret:2024wlu, Rignon-Bret:2024mef} for the relationship between covariance and central charges). That is, non-covariance/anomaly in the charge algebra manifests through the central charges in the algebra. 

We close this subsection by noting that, as mentioned, a covariant $\bar{Y}$-freedom does not alter the algebra and therefore does not affect the central extension term. However, non-covariant freedom can lead to additional central extension terms. See \cite{Compere:2015bca} for examples of how a non-covariant $Y$-term can change the central charge. 
\subsection{Conservation of  surface charges} \label{sec:conservation-charge}

In this subsection, we show that while  surface charges \eqref{surface-charge-integrable} are integrable they are not conserved. We start with evaluating the difference of charges at two (partial) Cauchy $\cauchy_1, \cauchy_2$ which intersect the boundary $\bdry$ at codimension-two surfaces $\mathcal{S}_1, \mathcal{S}_2$ (see Fig.~\ref{fig:3dconservation}):
\begin{equation}
   \begin{split}
        Q_2-Q_1 & =\int_{\mathcal{S}_2}\d{}x_{\mu\nu}\, Q^{\mu\nu}_{\text{\tiny{N}}}(\xi)-\int_{\mathcal{S}_1}\d{}x_{\mu\nu}\, Q^{\mu\nu}_{\text{\tiny{N}}}(\xi)\\
        & =\int_{\mathcal{B}_{12}}\d{}x_{\mu}\, \partial_{\nu}Q_{\text{\tiny{N}}}^{\mu\nu}(\xi)= \int_{\mathcal{B}_{12}}\d{}x_{\mu}\, J_{\xi}^{\mu}
        =\int_{\mathcal{B}_{12}}\d{}x_{\mu}\, (\Theta_{\text{\tiny{LW}}}^{\mu}[\delta_{\xi}\Phi;\Phi]-\xi^{\mu}\, \mathcal{L})\\
        &=\int_{\mathcal{B}_{12}}\d{}x_{\mu}\, (\Theta_{\text{\tiny{LW}}}^{\mu}[\delta_{\xi}\Phi;\Phi]+\chi^{\mu}[\delta_{\xi}\Phi;\Phi]\, \mathcal{L})
        =\int_{\mathcal{B}_{12}}\d{}x_{\mu}\, \Theta^{\mu}[\delta_{\xi}\Phi;\Phi]\, ,
   \end{split}
\end{equation}
where $\bdry_{12}$ is a segment of $\bdry$ between $\cauchy_2$ and $\cauchy_1$. The apparent non-conservation of the charge should not come as a surprise because we allow for our boundaries $\bdry, \cauchy$ (and hence ${\cal S}$) to have non-zero fluctuations. This in particular means we are using different time coordinates (and derivatives) at the two Cauchy surfaces ${\cal C}_1, {\cal C}_2$; the non-conservation is quantifying the fluctuation in the time units.
\section{Discussion}\label{sec:conc}
In this work, we extended CPSF to the cases where the boundaries are allowed to fluctuate. This naturally led to a focus on the codimension-two ``corner'' setup (in contrast to a codimension-one boundary setting). We studied how boundary fluctuations modify the symplectic two-form and surface charge associated with diffeomorphisms. As we established modifications in the surface charges are such that they always become integrable and that  the Noether charge is the integrable charge. We also derived the charge algebra and demonstrated how central charges  emerge in the algebra due to non-covariance/anomaly of the ``reference point'' of the charge. (The non-covariance/anomaly in a quantity $X$ is defined in \eqref{def-covariant}).

We  discussed that the two notions of (non)integrability and (non)conservation of the charges which are closely related in fixed-boundary analyses, become distinct when we allow boundaries to also fluctuate. This happens because the notion of (non)conservation comes from comparing values of a given charge at to different (partial) Cauchy surfaces ${\cal C}_1, {\cal C}_2$. If ${\cal C}$ is allowed to fluctuate, conservation does not imply equal value of charges on  ${\cal C}_1, {\cal C}_2$. In the analysis of  the symplectic form, this implies that the symplectic flux through ${\cal C}_1, {\cal C}_2$ should be balanced by the flux through the boundary ${\cal B}_{12}$, cf. Fig.~\ref{fig:3dconservation} and \eqref{symp-form-cons}.

Our analyses and results may be viewed as complementary to those in \cite{Ciambelli:2021nmv, Freidel:2021dxw, Carrozza:2022xut}. The main object in constructions of \cite{Ciambelli:2021nmv, Carrozza:2022xut} is the $(1,1)$-form $\chi^*_\mu$. Our ``fluctuating boundary'' setup allowed us to explicitly construct $\chi^*_\mu$ in terms of spacetime and field space variations of the boundary fields ${\cal A}, {\cal B}, {\cal C}$ \eqref{chi*-def}. Note also the presence of a fundamental object like $\chi^*_\mu$ which is a one-form in the 2-dim. part of spacetime transverse to the corner ${\cal S}$ breaks the $sl(2,\mathbb{R})$ ``corner symmetry'' (diffeomorphisms) that rotates us in this 2-dim. part of spacetime. Our formulation clarifies this feature too: in our setting ${\cal E}, \mathbb{S}$ that are  $sl(2,\mathbb{R})$ covariant quantities, are fundamental objects and  $\chi^*_\mu$ is specified in terms of these through \eqref{E-S-chi}. While these equations are $sl(2,\mathbb{R})$ covariant any specific solution to them breaks this symmetry. Explicitly, these equations specify $\chi^*_\mu$ up to spacetime or field space exact forms, up to $(1,0)$ and $(0,1)$ forms, and any specific choice for these ``integration constants'' breaks the symmetry. The situation here is analogous to Yang-Mills gauge theory field equations, which may be written in terms of gauge covariant quantity, gauge field strength. However, any solution for the one-form gauge field breaks the gauge invariance. Here $\chi^*_\mu$ is analogous to the ``Yang-Mills gauge field'' which is a one-form over the spacetime and a one-form in the field space (playing the role of the fundamental representation of the gauge algebra in our analogy). In this analogy \eqref{E-S-chi} may be viewed as the ``gauge field equations''.  While this analogy may prompt viewing \eqref{E-S-chi} as the Maurer-Cartan equation (as suggested in \cite{Carrozza:2022xut} or in eq.(14) of \cite{Ciambelli:2021nmv}), the explicit form of \eqref{E-S-chi} does not show this, unless one has an equation which relates two-forms over the field space to one-forms there; in the analogy with Yang-Mills theory the gauge algebra does this. To study this point one should have an explicit construction of the solution space and study the charges, which we intend to do in future work.

In this work, we mainly developed the basic mathematical formulation of covariant phase space formalism with fluctuating boundaries. Here we briefly discuss the physical applications/implications of this formulation. These deserve a thorough study which is postponed to future publications. 

\paragraph{Black hole thermodynamics.} The seminal work of Iyer and Wald \cite{Iyer:1994ys} established that standard CPSF  for bifurcate Killing horizons provides a robust proof for the first law of black hole thermodynamics for any diffeomorphism invariant theories of gravity. This proof may be extended to a generalized/local version of zeroth and first laws once we apply CPSF to fix null boundaries \cite{Adami:2021kvx}. It would be interesting to revisit and extend the analysis of \cite{Adami:2021kvx} to the ``boundary fluctuating'' setup developed in this work. Besides the zeroth and first law, it is desirable to also explore the second law in this setting. 

\paragraph{Dynamical black holes and information problem.} As pointed out our setup is a codimension-two corner-based one (in contrast to the codimension-one boundary-based one). This setup is particularly interesting and useful to analyze the dynamics of black holes once we view our corner ${\cal S}$ as the bifurcation surface of a black hole. In the boundary-based setup \cite{Adami:2020amw, Adami:2021nnf} one can't naturally analyze the bifurcation point (which is the infinite coordinate limit on the null boundary). On the other hand to study dynamical black holes one should necessarily use a formulation in which the bifurcation surface is also allowed to be dynamical and fluctuate. We hope our formulation here not only generalizes and extends the discussion in  \cite{Hollands:2024vbe} for dynamical horizons but also provides the setup to address the information problem.

\paragraph{Memory effects.} We close with a comment on (gravitational) memory effects, permanent changes in a physical observable of a detector (like its position, velocity, spin, ...) due to the passage of gravitational waves \cite{Zeldovich:1974gvh, Braginsky:1985vlg, Braginsky:1987kwo, Thorne:1992sdb, Favata:2010zu}. Memory effects have been discussed to be related to the changes in the ``soft'' charges (surface charges) \cite{Strominger:2017zoo, Compere:2018ylh, Pasterski:2021rjz}. In other words, the non-conservation of the charges is recorded in memory effects. As pointed out, the non-integrability of surface charges is related to the flux of the associated physical observable due to the genuine flux of gravitational waves, see \cite{Adami:2021nnf}. On the other hand,  non-conservation of surface charges is closely related to non-integrability of the charges in a fixed timelike or null boundary through the charge-flux balance equations \cite{Barnich:2011mi}. The latter in gravitational theories are equations like Raychaudhuri (or Bondi) or Damour equations \cite{Adami:2021nnf}. As we discussed when we allow for boundaries to fluctuate, non-conservation and non-integrability are not related to each other as in the fixed-boundary cases. It would be instructive to study and clarify this in more detail.

\section*{Acknowledgement}
We would like to thank Luca Ciambelli for fruitful discussions. 
The work of HA is supported
by Beijing Natural Science Foundation under Grant No IS23018. The work of MMShJ is supported in part by  Iran National Science Foundation (INSF) Grant No.  4026712 and in part the ICTP through the senior Associates Programme (2023-2028) and also funds from ICTP HECAP section. {The work of MG is partially supported by IPM funds.}

\appendix

\section{Useful identities of distributions}\label{sec:identity}
Given two scalar functions ${\cal B}, {\cal C}$ and the step-function  $H(X)$ and Dirac delta-function $\Delta(X)=H'(X)$, one has the following identities
\begin{equation}
    \delta H(X)=\Delta(X)\, \delta \bdry\, , \qquad \delta\Delta(X)=\Delta'(X) \, \delta X\, .
\end{equation}
\begin{equation}\label{Delta_B_C}
  \Delta(\bdry)=- N_{\bdry}\, b^{\mu}\, \partial_{\mu} H(\bdry)\, , \qquad \Delta(\cauchy)= N\, n^{\mu}\, \partial_{\mu} H(\cauchy)\, .
\end{equation}
\begin{equation}\label{DeltaPrime_B_C}
  \Delta'(\bdry)=- N_{\bdry}\, b^{\mu}\, \partial_{\mu}\Delta(\bdry)\, , \qquad     \Delta'(\cauchy)= N\, n^{\mu}\, \partial_{\mu}\Delta(\cauchy)\, .
\end{equation}

\section{Variation of  integrals}\label{Appen:variations}
In this appendix we present some identities for variations of codimension-0, codimension-one, and codimension-two integrals.

\paragraph{Integrals of $(D,0)$-forms.} Assume $X$ is a scalar density in the spacetime and a 0-form in the phase space. The variation of a codimension-0 integral is as follows
\begin{equation}
    \begin{split}
        \delta \int_{\MBL}\, \d{}^{D}x\, X\, &=\delta \int_{\mathcal{M}}\, \d{}^{D}x\, X\, H(\bdry)=\int_{\mathcal{M}}\, \d{}^{D}x\, [\delta X\, H(\bdry)+X\, \delta H(\bdry)]\\
        &=\int_{\mathcal{M}}\, \d{}^{D}x\, [\delta X\, H(\bdry)+X\, \Delta(\bdry)\, \delta \bdry]
        =\int_{\mathcal{M}} \, \d{}^{D}x\,\delta X\,  H(\bdry)+\int_{\bdry} \, \d{}^{D-1}x X\, \delta \bdry\, .
    \end{split}
\end{equation}
Hence we get the following identity
\begin{equation}\label{id-D-0}
    \boxed{\delta \int_{\MBL}\, \d{}^{D}x\,X =\int_{\mathcal{M}}\d{}^{D}x\, \delta X\,  H(\bdry)+\int_{\bdry} \d{}x_{\mu}\, \chi^{\mu}\, X\, .}
\end{equation}

\paragraph{Integrals of $(D-1,0)$-forms.} Consider a vector  density $X^\mu$ (dual to $D-1$-form) in spacetime and a 0-form in solution space, such that $\d {} x_\mu\, X^\mu= \d {}^{D-1} x\, X$. Then
\begin{align}
        \delta \int_{\Bplus} \d {}^{D-1} x\, X\, &=\delta \int_{\mathcal{M}} \d {}^{D} x\, X\, H(\cauchy)\, \Delta (\bdry) \nonumber\\
        &=\int_{\mathcal{M}} \d {}^{D} x\, \Big[\delta X\, H(\cauchy)\, \Delta (\bdry)+ X\, \delta H(\cauchy)\, \Delta (\bdry)+  X\, H(\cauchy)\, \delta\Delta (\bdry)\Big] \nonumber\\
        &=\int_{\mathcal{M}} \d {}^{D} x\, \Big[\delta X\, H(\cauchy)\, \Delta (\bdry)+ X\, \Delta(\cauchy)\, \delta \cauchy\, \Delta (\bdry)+  X\, H(\cauchy)\, \Delta'(\bdry)\, \delta \bdry\Big] \nonumber\\
        &=\int_{\mathcal{M}} \d {}^{D} x\, \Big[\delta X\, H(\cauchy)\, \Delta (\bdry) + X\, N\, n^{\mu}\, \partial_{\mu} H(\cauchy)\, \delta \cauchy\, \Delta (\bdry)- X\, N_{\bdry}\, {b^{\mu}}\, \partial_{\mu}\Delta(\bdry) \, H(\cauchy)\, \delta \bdry\big]\nonumber\\
        &=\int_{\mathcal{M}} \d {}^{D} x\, \Big[\delta X\, H(\cauchy)\, \Delta (\bdry) -\partial_{\mu}(X\, N\, {n^{\mu}}\, \delta \cauchy)H(\cauchy)\, \Delta (\bdry) + \partial_{\mu}( X\, N_{\bdry}\, {b^{\mu}}\, \delta \bdry)H(\cauchy)\, \Delta (\bdry)\nonumber\\
        &\hspace{2 cm}-X\, N_{\bdry}\, N_{\cauchy}^{-1} b^{\mu}\, \delta \bdry \, c_{\mu}\, \Delta(\bdry)\,\Delta(\cauchy)\Big]\nonumber\\
        &=\int_{\mathcal{M}} \d {}^{D} x\, \Big[\delta X\, H(\cauchy)\, \Delta (\bdry) -\partial_{\mu}(X\, N\, {n^{\mu}}\, \delta \cauchy)H(\cauchy)\, \Delta (\bdry) + \partial_{\mu}( X\, N_{\bdry}\, {b^{\mu}}\, \delta \bdry)H(\cauchy)\, \Delta (\bdry)\nonumber\\
        &\hspace{2 cm}-X\, N_{\bdry}\, b^{\mu}\, \delta \bdry (N^{-1}\, n_{\mu}+{\cal A}\, N^{-1}_{\bdry}\, b_{\mu})\, \Delta(\bdry)\,\Delta(\cauchy)\Big]\nonumber\\
        &=\int_{\mathcal{M}} \d {}^{D} x\, \Big[\delta X\, H(\cauchy)\, \Delta (\bdry) -\partial_{\mu}(X\, N\, {n^{\mu}}\, \delta \cauchy)H(\cauchy)\, \Delta (\bdry) + \partial_{\mu}( X\, N_{\bdry}\, {b^{\mu}}\, \delta \bdry)H(\cauchy)\, \Delta (\bdry)\nonumber\\
        &\hspace{2 cm}-X\, \mathcal{A}\, \delta \bdry\, \Delta(\bdry)\,\Delta(\cauchy)\Big]\nonumber\\
        &=\int_{\mathcal{M}} \d {}^{D} x\, \Big[\delta X\, H(\cauchy)\, \Delta (\bdry) -\partial_{\mu}(X\, N\, {n^{\mu}}\, \delta \cauchy)H(\cauchy)\, \Delta (\bdry) + \partial_{\mu}( X\, N_{\bdry}\, {b^{\mu}}\, \delta \bdry)H(\cauchy)\, \Delta (\bdry)\nonumber\\
        &\hspace{2 cm}-X\, \mathcal{A}\, \delta \bdry\, \Delta(\bdry)\, N\, n^{\mu}\, \partial_{\mu} H(\cauchy)\Big]\nonumber\\
        &=\int_{\mathcal{M}} \d {}^{D} x\, \Big[\delta X\, H(\cauchy)\, \Delta (\bdry) -\partial_{\mu}(X\, N\, {n^{\mu}}\, \delta \cauchy)H(\cauchy)\, \Delta (\bdry) + \partial_{\mu}( X\, N_{\bdry}\, {b^{\mu}}\, \delta \bdry)H(\cauchy)\, \Delta (\bdry)\nonumber\\
        &\hspace{2 cm}+\partial_{\mu}(X\, \mathcal{A}\, \delta \bdry\,  N\, n^{\mu}\,  )H(\cauchy)\Delta(\bdry)\,\Big]\nonumber\\
        &=\int_{\mathcal{M}} \d {}^{D} x\, \Big[\delta X +\partial_{\mu}(-X\, N\, {n^{\mu}}\, \delta \cauchy+ X\, N_{\bdry}\, {b^{\mu}}\, \delta \bdry+X\, \mathcal{A}\, \delta \bdry\, N\, n^{\mu})\Big]H(\cauchy)\, \Delta (\bdry)\, ,\nonumber
\end{align}
where we have used \eqref{Delta_B_C},  and $b_{\mu}\, n^{\mu}=0$ and also $\chi$ is given in \eqref{chi-def}. So, we have the following useful identity
\begin{equation}\label{identity-1}
    \boxed{\delta \int_{\Bplus} \d {}^{D-1} x\, X =\int_{\bdry} \d {}^{D-1} x\, \Big(\delta X+\partial_{\mu}(X\, \chi^{\mu})\Big)H(\cauchy)\, .}
\end{equation}

\paragraph{Integrals of $(D-1,1)$-form.} If the vector density $X^{\mu}$ in the above is one-form in solution space, repeat the analysis as in the previous case and arrive at,
\begin{equation}\label{codim1-1form}
    \boxed{\delta\int_{\Bplus} \d{}x_{\mu}\, X^{\mu} =\int_{\bdry} \d{}x_{\mu}\,\Big( \delta X^{\mu} \, + \chi^{\mu} \wedge \partial_{\nu} X^{\nu} \Big) H(\cauchy)+2\int_{{\cal S}}\d{}x_{\mu\nu}\, X^{\mu}  \wedge \chi^{\nu}\, .
    }
\end{equation}

\paragraph{Integrals of $(D-2,0)$-form.} Let $X^{\mu\nu}$ be an antisymmetric bi-vector density, and define
$ \d{}x_{\mu\nu} X^{\mu\nu}:= \d{}^{D-2}x X$ where $X$ is a scalar density in spacetime and a 0-form in phase space. Then, 
\begin{align*}
 \hspace*{-5mm}   \delta \int_{{\cal S}} \d{}x_{\mu\nu} X^{\mu\nu}&= \, \delta \int_{{\cal S}} \d{}^{D-2}x\, X = \delta \int_{\mathcal{M}} \d{}^{D}x\, X\, \Delta (\bdry)\, \Delta(\cauchy)\\
    &=\int_{\mathcal{M}} \d{}^{D}x\, [\delta X\, \Delta (\bdry)\, \Delta(\cauchy)+X\, \delta \Delta (\bdry)\, \Delta(\cauchy)+X\, \Delta (\bdry)\, \delta\Delta(\cauchy)]\\
    &=\int_{\mathcal{M}} \d{}^{D}x\, \Big[\delta X\, \Delta (\bdry)\, \Delta(\cauchy)+X\,  \Delta' (\bdry)\, \delta \bdry\, \Delta(\cauchy)+X\, \Delta (\bdry)\, \Delta' (\cauchy)\, \delta \cauchy \Big]\\
    &=\int_{\mathcal{M}} \d{}^{D}x\, \Big[\delta X\, \Delta (\bdry)\, \Delta(\cauchy)+X (- N_{\bdry}\, b^{\mu}\, \partial_{\mu}\Delta(\bdry)) \delta \bdry\, \Delta(\cauchy)+X\, \Delta (\bdry)(N\, n^{\mu}\, \partial_{\mu}\Delta(\cauchy)) \delta \cauchy \\
    &=\int_{\mathcal{M}} \d{}^{D}x\, \Big[\delta X\, \Delta (\bdry)\, \Delta(\cauchy)+\partial_{\mu}(X\,  N_{\bdry}\, b^{\mu}\,  \delta \bdry) \Delta(\bdry)\Delta(\cauchy)+X\,  N_{\bdry}\, b^{\mu}\,  \delta \bdry \Delta(\bdry)\Delta'(\cauchy)\partial_{\mu}\cauchy\\
    &\hspace{2 cm}-\partial_{\mu} (X\, N \, n^{\mu}\, \delta \cauchy) \Delta(\cauchy)\Delta (\bdry) \Big]\\
    &=\int_{\mathcal{M}} \d{}^{D}x\, \Big[\delta X\, \Delta (\bdry)\, \Delta(\cauchy)+\partial_{\mu}(X\,  N_{\bdry}\, b^{\mu}\,  \delta \bdry) \Delta(\bdry)\Delta(\cauchy)-X\, \mathcal{A}\,   \delta \bdry \Delta(\bdry)\Delta'(\cauchy)\\
    &\hspace{2 cm}-\partial_{\mu} (X\, N \, n^{\mu}\, \delta \cauchy) \Delta(\cauchy)\Delta (\bdry) \Big]\\
    &=\int_{\mathcal{M}} \d{}^{D}x\, \Big[\delta X\, \Delta (\bdry)\, \Delta(\cauchy)+\partial_{\mu}(X\,  N_{\bdry}\, b^{\mu}\,  \delta \bdry) \Delta(\bdry)\Delta(\cauchy)-X\, N\, n^{\mu}\,  \mathcal{A}\,   \delta \bdry \Delta(\bdry)\, \partial_{\mu}\Delta(\cauchy)\\
    &\hspace{2 cm}-\partial_{\mu} (X\, N \, n^{\mu}\, \delta \cauchy) \Delta(\cauchy)\Delta (\bdry) \Big]\\
    &=\int_{\mathcal{M}} \d{}^{D}x\, \Big[\delta X\, \Delta (\bdry)\, \Delta(\cauchy)+\partial_{\mu}(X\,  N_{\bdry}\, b^{\mu}\,  \delta \bdry) \Delta(\bdry)\Delta(\cauchy)+\partial_{\mu}(X\, N\, n^{\mu}\,  \mathcal{A}\,   \delta \bdry) \Delta(\bdry)\, \Delta(\cauchy)\\
    &\hspace{2 cm}-\partial_{\mu} (X\, N \, n^{\mu}\, \delta \cauchy) \Delta(\cauchy)\Delta (\bdry) \Big]\\
    &=\int_{\mathcal{M}} \d {}^{D} x\, \Big[\delta X +\partial_{\mu}(-X\, N\, {n^{\mu}}\, \delta \cauchy+ X\, N_{\bdry}\, {b^{\mu}}\, \delta \bdry+X\,N\, n^{\mu} \mathcal{A}\, \delta \bdry) \Big]\Delta(\cauchy)\, \Delta (\bdry)\, .
\end{align*}
Finally, we reach the following identity
\begin{equation}\label{identity-2}
    \boxed{\delta \int_{\cal S}  \d{} x_{\mu\nu}\ X^{\mu\nu}=\int_{\cal S} \d{}^{D-2}x\, \Big(\delta X+\partial_{\mu}(X\, \chi^{\mu})\Big)\, .}
\end{equation}
\paragraph{Integrals of $(D-2,1)$-form.}
Let $X^{\mu\nu}$ be an antisymmetric bi-vector density in spacetime and a one-form in phase space. Then, one can repeat the analysis as in the previous cases and arrive at 
\begin{equation}\label{identity-delta-2-1}
    \boxed{\delta \int_{{\cal S}}\d{}x_{\mu\nu}\, X^{\mu\nu}=\int_{{\cal S}}\d{}x_{\mu\nu}\, (\delta X^{\mu\nu}-2\partial_{\alpha} X^{\mu\alpha}\wedge \chi^{\nu})\, .}
\end{equation}

\section{Details of calculations}
\paragraph{$W$-freedom in pre-symplectic potential.}
By considering $W$-freedom, the action \eqref{action-original} is as follows
\begin{equation}
\begin{split}
    \tilde{S}=S+\int_{\MBL} \d{}^{D}x\,  \partial_\mu W^\mu &=S+\int_{\mathcal{M}} \d{}^{D}x\,  \partial_\mu W^\mu\,  H(\bdry)\, .
\end{split}
\end{equation}
Now let us compute its variation. From identity \eqref{id-D-0} we have
\begin{equation}
    \begin{split}
    \delta \tilde{S}&=\delta S+\int_{\mathcal{M}} \d{}^{D}x\,  \partial_\mu \delta W^\mu\,  H(\bdry)+\int_{\mathcal{\bdry}} \d{} x_\mu \,\chi^\mu\, \partial_\nu W^\nu\, \\
    &=\delta S+\int_{\mathcal{\bdry}} \d{} x_\mu \,( \delta W^\mu + \chi^\mu\, \partial_\nu W^\nu)\, .
    \end{split}
\end{equation}
So, due to the $W$-freedom, the symplectic potential transforms as follows
\begin{equation}
    \tilde{\Theta}^\mu = \Theta^\mu + \chi^\mu \mathcal{L}+\delta W^\mu+\chi^\mu \partial_\nu W^\nu\, .
\end{equation}

\paragraph{Proof of the identity:}
\begin{equation}\label{E-del-chi}
    \begin{split}
        \boxed{\mathcal{E}_{\mu \nu}(\partial_{\alpha}\chi^{\mu} \wedge \chi^{\alpha}-\delta\chi^{\mu} )=0\, .}
    \end{split}
\end{equation}
We start from \eqref{chi*-def} and \eqref{basic-form-relation-2}
\begin{equation}
\partial_\mu \mathbb{S} ={-}\delta \chi^*_{\mu}\, , \qquad \quad \mathbb{S}=\frac{1}{2}\chi^{\mu}\wedge \chi^*_{\mu}\,  ,\qquad \chi^*_{\mu}= \mathcal{E}_{\mu \nu} \chi^\nu\, .
\end{equation}
First, we consider the left-hand side of the above equation
\begin{equation}
\begin{split}
\partial_\alpha \mathbb{S}&=\frac{1}{2}\partial_\alpha(\chi^{\mu}\wedge \chi^*_{\mu})\, =\frac{1}{2}\partial_\alpha(\mathcal{E}_{\mu \nu}\chi^{\mu}\wedge \chi^{\nu})\, \\
&=\frac{1}{2}\Big(\partial_\alpha \mathcal{E}_{\mu \nu}\,\chi^{\mu}\wedge \chi^{\nu}+\mathcal{E}_{\mu \nu}\partial_\alpha\chi^{\mu}\wedge \chi^{\nu}+\mathcal{E}_{\mu \nu}\chi^{\mu}\wedge \partial_\alpha \chi^{\nu}\Big)\,\\
&=\frac{1}{2}\partial_\alpha \mathcal{E}_{\mu \nu}\,\chi^{\mu}\wedge \chi^{\nu}+\mathcal{E}_{\mu \nu}\partial_\alpha\chi^{\mu}\wedge \chi^{\nu}\, .
\end{split}
\end{equation}
Now let us look at the right-hand side    
\begin{align*}
-\delta \chi^*_{\alpha}&=- \delta \left(\mathcal{E}_{\alpha \nu} \chi^\nu\right)=- \delta \mathcal{E}_{\alpha \nu} \wedge \chi^\nu- \mathcal{E}_{\alpha \nu} \,  \delta \chi^\nu \\
&=- 2 \partial_{[\alpha} \chi^*_{\nu]}\,\wedge \chi^\nu- \mathcal{E}_{\alpha \nu} \,  \delta \chi^\nu
= (\partial_{\nu} \chi^*_{\alpha}-\partial_{\alpha} \chi^*_{\nu})\wedge \chi^\nu- \mathcal{E}_{\alpha \nu} \,  \delta \chi^\nu\\
&= (\partial_{\nu} (\mathcal{E}_{\alpha \mu} \chi^\mu)-\partial_{\alpha} (\mathcal{E}_{\nu \mu}\chi^\mu))\wedge \chi^\nu- \mathcal{E}_{\alpha \nu} \,  \delta \chi^\nu\\
&= \left(\partial_{\nu} \mathcal{E}_{\alpha \mu}\, \chi^\mu+ \mathcal{E}_{\alpha \mu} \,\partial_{\nu}\chi^\mu-\partial_{\alpha} \mathcal{E}_{\nu \mu}\,\chi^\mu- \mathcal{E}_{\nu \mu}\,\partial_{\alpha}\chi^\mu\right)\wedge \chi^\nu- \mathcal{E}_{\alpha \nu} \,  \delta \chi^\nu\\
&= \partial_{\nu} \mathcal{E}_{\alpha \mu}\, \chi^\mu\wedge \chi^\nu+ \mathcal{E}_{\alpha \mu} \,\partial_{\nu}\chi^\mu\wedge \chi^\nu-\partial_{\alpha} \mathcal{E}_{\nu \mu}\,\chi^\mu\wedge \chi^\nu+ \mathcal{E}_{\mu\nu}\,\partial_{\alpha}\chi^\mu\wedge \chi^\nu- \mathcal{E}_{\alpha \nu} \,  \delta \chi^\nu \, .    
\end{align*}
So, considering both sides we have
\begin{equation}
    \begin{split}
   \mathcal{E}_{\alpha \mu} \,\partial_{\nu}\chi^\mu\wedge \chi^\nu- \mathcal{E}_{\alpha \nu} \,  \delta \chi^\nu  &=\frac{1}{2}\partial_\alpha \mathcal{E}_{\mu \nu}\,\chi^{\mu}\wedge \chi^{\nu}- \partial_{\nu} \mathcal{E}_{\alpha \mu}\, \chi^\mu\wedge \chi^\nu+\partial_{\alpha} \mathcal{E}_{\nu \mu}\,\chi^\mu\wedge \chi^\nu\\
   &=-\frac{1}{2}\partial_\alpha \mathcal{E}_{\mu \nu}\,\chi^{\mu}\wedge \chi^{\nu}- \partial_{\nu} \mathcal{E}_{\alpha \mu}\, \chi^\mu\wedge \chi^\nu\\
   &=-\frac{1}{2}\partial_\alpha \mathcal{E}_{\mu \nu}\,\chi^{\mu}\wedge \chi^{\nu}-\frac{1}{2} \partial_{\nu} \mathcal{E}_{\alpha \mu}\, \chi^\mu\wedge \chi^\nu+\frac{1}{2} \partial_{\nu} \mathcal{E}_{\mu\alpha}\, \chi^\mu\wedge \chi^\nu\\
   &=-\frac{1}{2}\partial_\alpha \mathcal{E}_{\mu \nu}\,\chi^{\mu}\wedge \chi^{\nu}-\frac{1}{2} \partial_{\nu} \mathcal{E}_{\alpha \mu}\, \chi^\mu\wedge \chi^\nu-\frac{1}{2} \partial_{\mu} \mathcal{E}_{\nu\alpha}\, \chi^\mu\wedge \chi^\nu\\
   &=-\frac{3}{2}\partial_{[\alpha} \mathcal{E}_{\mu \nu]}\,\chi^{\mu}\wedge \chi^{\nu}=0 \, .\nonumber
    \end{split}
\end{equation}
In the last line, we used $\partial_{[\alpha}{\cal E}_{\mu\nu]}=\partial_{[\alpha} (\partial_\mu{\cal C} \partial_{\nu]}{\cal B})=0$.

By contracting both sides of identity \eqref{E-del-chi} with $n^\nu$ and $b^\nu$, immediately we get
\begin{equation}\label{b-del-chi}
    \begin{split}
        b_\mu (\partial_{\nu}\chi^{\mu} \wedge \chi^{\nu}-\delta\chi^{\mu})=0\, ,\qquad 
        n_\mu (\partial_{\nu}\chi^{\mu} \wedge \chi^{\nu}-\delta\chi^{\mu})=0\, .
    \end{split}
\end{equation}

\paragraph{Proof of the identity:} 
\begin{equation}\label{identity-E-chi}
\boxed{
\mathcal{E}_{\mu \nu} \chi^{\alpha}=2 q^\alpha_{[\mu} \chi^*_{\nu]} -2 \delta^\alpha_{[\mu} \chi^*_{\nu]\, .} 
}
\end{equation}
We start from \eqref{bi-normal} and \eqref{chi-def},
\begin{equation}
\begin{split}
    {\cal E}_{\mu\nu}\, \chi^{\alpha}&= N^{-1} N_{\bdry}^{-1} \qty(n_{\mu} b_{\nu} - n_{\nu} b_{\mu})\qty(-N\, {n^{\alpha}}\, \delta \cauchy + N_{\bdry}\, {b^{\alpha}}\, \delta \bdry+ N\, \mathcal{A}\, n^{\alpha}\, \delta \bdry)\\
    &=  \qty(n_{\mu} b_{\nu} - n_{\nu} b_{\mu})\qty(-N_{\bdry}^{-1}\, {n^{\alpha}}\, \delta \cauchy + N^{-1}\, {b^{\alpha}}\, \delta \bdry+ N_\bdry^{-1}\, \mathcal{A}\, n^{\alpha}\, \delta \bdry)\\
    &= -N_\bdry^{-1}\, n_{\mu} b_{\nu} \, n^{\alpha}\,( \delta \cauchy-\mathcal{A}\,\delta \bdry )+N^{-1}\,   n_{\mu} b_{\nu} \, b^{\alpha}\, \delta \bdry- (\mu \leftrightarrow \nu)\\
    &= N_\bdry^{-1}\, b_{\nu} \, (\delta^\alpha_\mu -q^\alpha_\mu -b^\alpha b_{\mu})\, ( \delta \cauchy-\mathcal{A}\,\delta \bdry ) +N^{-1}\,   n_{\mu} (\delta^\alpha_\nu -q^\alpha_\nu +n^\alpha n_{\nu})\, \delta \bdry- (\mu \leftrightarrow \nu)\\
    &= N_\bdry^{-1}\, b_{\nu} \, (\delta^\alpha_\mu -q^\alpha_\mu)\, ( \delta \cauchy-\mathcal{A}\,\delta \bdry ) +N^{-1}\,   n_{\mu} (\delta^\alpha_\nu -q^\alpha_\nu)\, \delta \bdry - (\mu \leftrightarrow \nu)\\
    &= -\partial_{\nu}\bdry \, (\delta^\alpha_\mu -q^\alpha_\mu)\, ( \delta \cauchy-\mathcal{A}\,\delta \bdry ) -(\partial_\mu \cauchy - \mathcal{A} \, \partial_\mu \bdry)(\delta^\alpha_\nu -q^\alpha_\nu)\, \delta \bdry - (\mu \leftrightarrow \nu)\\
    &= (\delta^\alpha_\mu -q^\alpha_\mu)\,\qty[ -\partial_{\nu}\bdry \,( \delta \cauchy-\mathcal{A}\,\delta \bdry ) +(\partial_\nu \cauchy - \mathcal{A} \, \partial_\nu \bdry) \delta \bdry] - (\mu \leftrightarrow \nu)\\
    &= (\delta^\alpha_\mu -q^\alpha_\mu)\,\qty[ -\partial_{\nu}\bdry \, \delta \cauchy+\partial_\nu \cauchy\delta \bdry] - (\mu \leftrightarrow \nu)= -(\delta^\alpha_\mu -q^\alpha_\mu)\,\chi^*_\nu- (\mu \leftrightarrow \nu)\\
    &=(\delta^\alpha_\nu -q^\alpha_\nu)\chi^*_\mu  -(\delta^\alpha_\mu -q^\alpha_\mu )\chi^*_\nu
   =2 q^\alpha_{[\mu} \chi^*_{\nu]} -2 \delta^\alpha_{[\mu} \chi^*_{\nu]}  \nonumber
\end{split}
\end{equation}
where we used \eqref{induced-Sigmac} and \eqref{chi*-def}.

\paragraph{Details of derivation of  \eqref{symp-form-freedom}.}
Let us look at the following corner term
\begin{allowdisplaybreaks}
	\begin{align}\label{Omega-freedom}
			&\int_{\mathcal{S}}\d{}x_{\mu\nu}\left[2\delta W^{\mu}\wedge \chi^{\nu}-\delta Y^{\mu\nu} +(2\partial_{\alpha}Y^{\mu\alpha}+\partial_{\alpha}W^{\alpha}\, \chi^{\mu})\wedge \chi^{\nu}\right]\\
			&=\int_{\mathcal{S}}\d{}x_{\mu\nu}\left[-\delta \bar{Y}^{\mu\nu}-2W^{\mu}\ \delta \chi^{\nu} +\qty(2\partial_{\alpha}\qty(\bar{Y}^{\mu\alpha}+2 W^{[\mu}\ \chi^{\alpha]})+\partial_{\alpha}W^{\alpha}\, \chi^{\mu})\wedge \chi^{\nu}\right]\nonumber \\
			&=\int_{\mathcal{S}}\d{}x_{\mu\nu}\left[-\delta \bar{Y}^{\mu\nu}+2\partial_{\alpha}\bar{Y}^{\mu\alpha}\wedge \chi^{\nu}-2W^{\mu}\ \delta \chi^{\nu} +\qty(2\partial_{\alpha}\qty( W^{\mu}\chi^{\alpha}- W^{\alpha}\chi^{\mu})+\partial_{\alpha}W^{\alpha}\, \chi^{\mu})\wedge \chi^{\nu}\right]\nonumber \\
			&=\int_{\mathcal{S}}\d{}x_{\mu\nu}\left[-\delta \bar{Y}^{\mu\nu}+2\partial_{\alpha}\bar{Y}^{\mu\alpha}\wedge \chi^{\nu}+2W^{\nu} \delta \chi^{\mu} +\qty(2\partial_{\alpha}\qty( W^{\mu}\chi^{\alpha})- 2 W^{\alpha}\partial_\alpha\chi^{\mu}-\partial_{\alpha}W^{\alpha}\, \chi^{\mu})\wedge \chi^{\nu}\right]\nonumber \\
			&=\int_{\mathcal{S}}\d{}x_{\mu\nu}\left[-\delta \bar{Y}^{\mu\nu}+2\partial_{\alpha}\bar{Y}^{\mu\alpha}\wedge \chi^{\nu}+2W^{\nu}\partial_{\alpha}\chi^{\mu} \wedge \chi^{\alpha} +\qty(2\partial_{\alpha}\qty( W^{\mu}\chi^{\alpha})- 2 W^{\alpha}\partial_\alpha\chi^{\mu}-\partial_{\alpha}W^{\alpha}\, \chi^{\mu})\wedge \chi^{\nu}\right]\nonumber \\
			&=\int_{\mathcal{S}}\d{}x_{\mu\nu}\left[-\delta \bar{Y}^{\mu\nu}+2\partial_{\alpha}\bar{Y}^{\mu\alpha}\wedge \chi^{\nu}+2\partial_{\alpha}(W^{\mu}\chi^{\alpha} \wedge \chi^{\nu}) -\qty(2 W^{\alpha}\partial_\alpha\chi^{\mu}+\partial_{\alpha}W^{\alpha}\, \chi^{\mu})\wedge \chi^{\nu}\right]\nonumber \\
			&=\int_{\mathcal{S}}\d{}x_{\mu\nu}\left[-\delta \bar{Y}^{\mu\nu}+2\partial_{\alpha}\bar{Y}^{\mu\alpha}\wedge \chi^{\nu}+2\partial_{\alpha}(W^{\mu}\chi^{\alpha} \wedge \chi^{\nu}) -\qty(2 \partial_\alpha (W^{\alpha}\chi^{\mu})-\partial_{\alpha}W^{\alpha}\, \chi^{\mu})\wedge \chi^{\nu}\right]\nonumber \\     
			&=\int_{\mathcal{S}}\d{}x_{\mu\nu}\left[-\delta \bar{Y}^{\mu\nu}+2\partial_{\alpha}\bar{Y}^{\mu\alpha}\wedge \chi^{\nu}+2\partial_{\alpha}(W^{\mu}\chi^{\alpha} \wedge \chi^{\nu}) -\qty(2 \partial_\alpha (W^{\alpha}\chi^{[\mu})\wedge \chi^{\nu]}-\partial_{\alpha}W^{\alpha}\, \chi^{\mu}\wedge \chi^{\nu})\right]\nonumber \\     
			&=\int_{\mathcal{S}}\d{}x_{\mu\nu}\left[-\delta \bar{Y}^{\mu\nu}+2\partial_{\alpha}\bar{Y}^{\mu\alpha}\wedge \chi^{\nu}+2\partial_{\alpha}(W^{\mu}\chi^{\alpha} \wedge \chi^{\nu}) -\qty( \partial_\alpha (W^{\alpha}\chi^{\mu})\wedge \chi^{\nu}-\partial_\alpha (W^{\alpha}\chi^{\nu})\wedge \chi^{\mu}-\partial_{\alpha}W^{\alpha}\, \chi^{\mu}\wedge \chi^{\nu})\right]\nonumber \\     
			&=\int_{\mathcal{S}}\d{}x_{\mu\nu}\Big[-\delta \bar{Y}^{\mu\nu}+2\partial_{\alpha}\bar{Y}^{\mu\alpha}\wedge \chi^{\nu}+2\partial_{\alpha}(W^{\mu}\chi^{\alpha} \wedge \chi^{\nu}) \nonumber \\
			&\hspace{20mm}-\qty( W^{\alpha}\partial_\alpha \chi^{\mu}\wedge \chi^{\nu}+\partial_\alpha W^{\alpha}\chi^{\mu}\wedge \chi^{\nu}-\partial_\alpha W^{\alpha}\chi^{\nu}\wedge \chi^{\mu}- W^{\alpha}\partial_\alpha\chi^{\nu}\wedge \chi^{\mu}-\partial_{\alpha}W^{\alpha}\, \chi^{\mu}\wedge \chi^{\nu})\Big]\nonumber \\     
			&=\int_{\mathcal{S}}\d{}x_{\mu\nu}\Big[-\delta \bar{Y}^{\mu\nu}+2\partial_{\alpha}\bar{Y}^{\mu\alpha}\wedge \chi^{\nu}+2\partial_{\alpha}(W^{\mu}\chi^{\alpha} \wedge \chi^{\nu}) \nonumber \\
			&\hspace{20mm}-\qty( W^{\alpha}\partial_\alpha \chi^{\mu}\wedge \chi^{\nu}+W^{\alpha} \chi^{\mu}\wedge\partial_\alpha\chi^{\nu}+\partial_{\alpha}W^{\alpha}\, \chi^{\mu}\wedge \chi^{\nu})\Big]\nonumber \\     
			&=\int_{\mathcal{S}}\d{}x_{\mu\nu}\Big[-\delta \bar{Y}^{\mu\nu}+2\partial_{\alpha}\bar{Y}^{\mu\alpha}\wedge \chi^{\nu}+2\partial_{\alpha}(W^{\mu}\chi^{\alpha} \wedge \chi^{\nu}) -\partial_\alpha \qty( W^{\alpha}\chi^{\mu}\wedge \chi^{\nu})\Big]\nonumber \\     
			&=\int_{\mathcal{S}}\d{}x_{\mu\nu}\Big[-\delta \bar{Y}^{\mu\nu}+2\partial_{\alpha}\bar{Y}^{\mu\alpha}\wedge \chi^{\nu}+\partial_{\alpha}(2W^{\mu}\chi^{\alpha} \wedge \chi^{\nu} - W^{\alpha}\chi^{\mu}\wedge \chi^{\nu})\Big]\nonumber \\
			&=\int_{\mathcal{S}}\d{}x_{\mu\nu}\Big[-\delta \bar{Y}^{\mu\nu}+2\partial_{\alpha}\bar{Y}^{\mu\alpha}\wedge \chi^{\nu}+{3}\partial_{\alpha}(W^{[\mu}\chi^{\alpha} \wedge \chi^{\nu]})\Big]\nonumber \\
			&=\int_{\mathcal{S}}\d{}x_{\mu\nu}\Big[-\delta \bar{Y}^{\mu\nu}+2\partial_{\alpha}\bar{Y}^{\mu\alpha}\wedge \chi^{\nu}\Big]+\text{total divergence term}\, .\nonumber
\end{align}\end{allowdisplaybreaks}
In the last line, we used $\partial_{[\alpha}{\cal E}_{\mu\nu]}=\partial_{[\alpha} (\partial_\mu{\cal C} \partial_{\nu]}{\cal B})=0$.
\bibliographystyle{fullsort.bst}
\bibliography{reference}


	\end{document}